\documentclass[12pt]{article}
\pdfoutput=1
\usepackage[utf8]{inputenc}
\usepackage{dsfont}
\usepackage{diagbox}
\usepackage{mathrsfs} 
\usepackage{amsfonts}
\usepackage{mathtools}
\allowdisplaybreaks[4]        
\usepackage{amssymb}
\usepackage{euscript}     
\usepackage{braket}
\usepackage{starfont}
\usepackage{color,soul}         
\usepackage{tensor}        
\usepackage{amsthm}
\usepackage{graphicx}
\usepackage{slashed}
\usepackage{leftidx}   
\usepackage{subfigure}
\usepackage{bbm}
\numberwithin{equation}{section}
\definecolor{outerspace}{rgb}{0.25, 0.29, 0.3}
\definecolor{scarlet}{rgb}{1.0, 0.13, 0.0}
\usepackage[header,title,page,titletoc]{appendix}  
\definecolor{princetonorange}{rgb}{1.0, 0.56, 0.0}
\definecolor{WildStrawberry}{rgb}{1.0, 0.26, 0.64}
\definecolor{rossocorsa}{rgb}{0.83, 0.0, 0.0}
\definecolor{navyblue}{rgb}{0.0, 0.0, 0.5}
\usepackage[numbers,sort&compress]{natbib}  
\usepackage{float}
\usepackage[paper=letterpaper,margin=1in]{geometry}
\newcommand{\req}[1]{(\ref{#1})} 

\newcommand{\eea}{\end{eqnarray}}
\newcommand{\ba}{\begin{eqnarray}}
\newcommand{\ea}{\end{eqnarray}}

\newcommand{\be}{\begin{equation}}
\newcommand{\ee}{\end{equation} }
\newcommand{\beqa}{\begin{eqnarray}}
\newcommand{\eeqa}{\end{eqnarray}}
\newcommand{\beqar}{\begin{eqnarray*}}
\newcommand{\eeqar}{\end{eqnarray*}}

\renewcommand{\req}[1]{eq.~(\ref{#1})}

\newcommand{\zp}{{\cal Z}_{+}}
\newcommand{\zpz}{{\cal Z}_{+}^{(0)}}
\newcommand{\zz}{{\cal Z}_{0}}
\newcommand{\zm}{{\cal Z}_{-}}
\newcommand{\zmz}{{\cal Z}_{-}^{(0)}}

\newcommand{\omegaz}{\omega^{(0)}}

\usepackage{hyperref}
\hypersetup{
    colorlinks,
    citecolor=rossocorsa,
    filecolor=navyblue,
    linkcolor=navyblue,
    urlcolor=navyblue
}

\begin{document} 

\begin{titlepage}

\begin{center}

\phantom{ }
\vspace{0cm}

{\bf \Large{On small black holes, KK monopoles\\
\vspace{0.25cm}
 and solitonic 5-branes}}
\vskip 0.5cm
Pablo A. Cano${}^{a}$, \'Angel Murcia${}^{b}$, Pedro F.~Ram\'{\i}rez${}^{c}$ and Alejandro Ruip\'erez${}^{d, e}$  
\vskip 0.05in

${}^{a}$\textit{Instituut voor Theoretische Fysica, KU Leuven \\
Celestijnenlaan 200D, B-3001 Leuven, Belgium}

\vskip -.2cm
${}^{b}$\textit{Instituto de F\'isica Te\'orica UAM/CSIC, \\
C/ Nicol\'as Cabrera, 13-15, C.U. Cantoblanco, 28049 Madrid, Spain}

\vskip -.2cm
${}^{c}$\textit{Max-Planck-Institut f\"ur Gravitationsphysik (Albert Einstein Institut), \\
Am M\"uhlenberg 1, D-14476 Potsdam, Germany}

\vskip -.2cm
${}^{d}$\textit{Dipartimento di Fisica ed Astronomia “Galileo Galilei”, Universit\`a di Padova, \\
Via Marzolo 8, 35131 Padova, Italy}

\vskip -.2cm
${}^{e}$\textit{INFN, Sezione di Padova, \\
Via Marzolo 8, 35131 Padova, Italy}

\begin{abstract}

We review and extend results on higher-curvature corrections to different configurations describing a superposition of heterotic strings, KK monopoles, solitonic 5-branes and momentum waves. Depending on which sources are present, the low-energy fields describe a black hole, a soliton or a naked singularity. We show that this property is unaltered when perturbative higher-curvature corrections are included, provided the sources are fixed. On the other hand, this character may be changed by appropriate introduction (or removal) of sources regardless of the presence of curvature corrections, which constitutes a non-perturbative modification of the departing system. The general system of multicenter KK monopoles and their 5-brane charge induced by higher-curvature corrections is discussed in some detail, with special attention paid to the possibility of merging monopoles. Our results are particularly relevant for small black holes (Dabholkar-Harvey states, DH), which remain singular after quadratic curvature corrections are taken into account. When there are four non-compact dimensions, we notice the existence of a black hole with regular horizon whose entropy coincides with that of the DH states, but the charges and supersymmetry preserved by both configurations are different. A similar construction with five non-compact dimensions is possible, in this case with the same charges as DH, although it fails to reproduce the DH entropy and supersymmetry. No such configuration exists if $d>5$, which we interpret as reflecting the necessity of having a 5-brane wrapping the compact space.

\end{abstract}
\end{center}

\end{titlepage}

\setcounter{tocdepth}{2}

{\parskip = .2\baselineskip \tableofcontents}

\section{Introduction}

Many supergravity theories are known to describe certain low energy limits of string theory. Hence, given a solution to the equations of motion of one of these supergravity theories, a natural question is to investigate if a correspondent description in terms of fundamental objects of string theory exists. Such a description is not always to be expected, as it is well-known that low-energy theories admit certain ``badly'' singular solutions which are to be regarded as unphysical (negative-mass Schwarzschild, for example) \cite{Horowitz:1995ta}. On the contrary, certain classical solutions, even if singular, can be argued to admit a microscopic interpretation provided some conditions are met --- see \cite{Gubser:2000nd, Mathur:2018tib}. In the case in which there are good reasons to expect a correspondent description, this identification turns out to pose a challenging problem unless the task is somehow facilitated. Simplifications take place when the system preserves some of the supersymmetries of the theory. From the field theory perspective, supersymmetry imposes relations between the components of the different fields, such that the allowed configurations are described by a reduced set of functions. Restrictions also occur for the equations of motion, with many of them being no longer independent. Typically, it suffices to solve Maxwell equations and Bianchi identities for some $p$-form fields, and it follows that the Einstein and scalar equations are automatically satisfied\footnote{For supersymmetric solutions with a null Killing vector, one component of the Einstein equations needs to be solved as well, as described for example in \cite{Gutowski:2003rg, Bellorin:2007yp, Cano:2018wnq}. It often occurs that a solution of this kind has appropriate isometries and can be equivalently described as a lower-dimensional configuration with a timelike Killing vector, in which case the original non-trivial Einstein equation is understood as a Maxwell or Bianchi equation of a $p$-form. The solutions described in this article have this property.} \cite{Kallosh:1993wx, Bellorin:2005hy}. Hence, one can say that a solution is completely determined by the specification of the charge distribution associated to the corresponding $p$-forms. On the UV part of the story, one then needs to find supersymmetric states in the spectrum acting as sources of those fields, an information that can be read from the worldsheet or worldbrane (effective) action. The identification obtained in this manner can be tested by comparing additional properties, like the number of supersymmetries preserved or the degeneracy. The use of these tools has been very fruitful, playing a role in much progress in string theory. Some noteworthy examples are the discovery of non-perturbative fundamental objects in the spectrum, evidence in favour of a web of dualities connecting seemingly distinct string theories or the identification of the microscopic degrees of freedom responsible for the thermodynamic entropy of certain black holes. A quite limited list of references is \cite{Buscher:1987sk, Strominger:1990et, Duff:1990wu, Callan:1991dj, Sen:1994fa, Schwarz:1994xn, Duff:1994zt, Polchinski:1995mt, Witten:1995ex, Polchinski:1996na, Strominger:1996sh, Maldacena:1996gb, Maldacena:1996ky, Kraus:2006wn, Sen:2007qy}. 

The microscopic derivation of black hole entropy performed by Strominger and Vafa followed a seminal paper of Sen that studied heterotic small black holes \cite{Sen:1996pb}, whose event horizon is singular and has zero size. Small black holes provide a toy model that was close to becoming the first confirmed description of black hole microstates in quantum gravity and, hence, their study has a special position in the history of the achievements of the theory. Consider states consisting of excitations of a string carrying winding and momentum charges $(\mathcal Q_w,\mathcal Q_n)$. This system was first studied by Dabholkar and Harvey (DH) in \cite{Dabholkar:1989jt} ---see also \cite{Dabholkar:1990yf}. In the heterotic theory, the degeneracy of these states gives the following value for the entropy in the large charge limit \cite{Russo:1994ev, Sen:1994eb}

\begin{equation}
\label{eq:smallentropy}
S=4 \pi \sqrt{\mathcal Q_n \mathcal Q_w} \, .
\end{equation} 

\noindent
The mass of the DH states grows linearly with the value of the charges. Hence, for large values of $(\mathcal Q_w, \mathcal Q_n)$ a black hole can be expected to emerge at the effective gravitational field theory \cite{Susskind:1993ws}. However, when one tries to construct such a black hole, a singular horizon with vanishing area is obtained and the formula \eqref{eq:smallentropy} is not reproduced. Sen argued that, since the effective theory shall not be valid in regions of large curvature, a ``stretched horizon" surface beyond which the usual understanding breaks down can be defined. He then postulated that the area of this stretched horizon would account for the macroscopic entropy of the system, and showed that the value, remarkably, scales with $\sqrt{\mathcal Q_n \mathcal Q_w}$.

Sen's insight found two lines of continuation. On the one hand, a string carrying momentum should oscillate, and one can study how many solutions can be constructed such that the string's profile lies within a stretched horizon \cite{Lunin:2002qf}. Depending on the duality frame used to describe them, the resulting geometries are of singular or solitonic nature. On the other hand, working within the special geometry formulation of effective four-dimensional supergravity with higher-curvature corrections, it was found in \cite{Dabholkar:2004yr, Dabholkar:2004dq} that it is possible to construct a regular near-horizon geometry reproducing \eqref{eq:smallentropy} such that only two of the lower dimensional vectors carry non-vanishing charge. The two approaches offer a distinct realization of the macroscopic entropy in the field theory, and a debate was opened regarding the compatibility of these two ideas \cite{Sen:2009bm, Mathur:2018tib}.

 In the light of the findings of \cite{Dabholkar:2004yr, Dabholkar:2004dq}, shortly followed by \cite{Sen:2004dp, Hubeny:2004ji}, it emerged the appealing idea that stringy higher-curvature corrections lead to the resolution of the singular small horizon.\footnote{See also \cite{Sen:2005kj, Sahoo:2006pm, Kraus:2006wn, Dabholkar:2006za, Sen:2007qy, Becker:2007zj, Prester:2008iu, Prester:2010cw, Polini:2019hda} and references therein.} String theory, as candidate to being a consistent theory of quantum gravity, is expected to resolve the singularities that mark the limitations of classical gravitational theories when these are associated to physically allowed configurations. For instance, one should be able to describe the collapse and evaporation of a black hole in terms of a unitary evolution free of divergences in a UV-complete theory. But the idea that stringy or quantum corrections may resolve singularities directly in the low-energy (field-theory) approximation goes beyond that expectation. It is, therefore, interesting to explore if this is actually a generic feature of the theory. Arguably, the simplest test that can be performed is to study similar configurations in slightly different situations. However, it turns out that the same mechanism that produced the horizon resolution in a few cases, failed in others without a clear explanation. Some examples of the latter case are those of a type II string with winding and momentum charges on a toroidal compactification \cite{Sen:2007qy}, or a heterotic string with five or more non-compact dimensions\footnote{A five-dimensional heterotic two charge solution with regular horizon exists, but its entropy differs from \eqref{eq:smallentropy}. We will discuss this solution in more detail in section~\ref{sec:fakesmall}.} \cite{Prester:2008iu, Sen:2005kj}. In view of these facts, it is fair to acknowledge that the effect of higher-curvature corrections must be understood better. In this article we study the problem by revisiting the original small-black-hole system directly in the original ten-dimensional heterotic theory, instead of using four-dimensional supergravity formulated in the language of special geometry as was done in \cite{Dabholkar:2004yr, Dabholkar:2004dq, Sen:2004dp, Hubeny:2004ji}. While we will not have at our disposal the powerful tools based on the attractor mechanism developed in \cite{Behrndt:1998eq, LopesCardoso:1998tkj, LopesCardoso:1999cv, LopesCardoso:1999fsj}, in exchange we will have analytic solutions in the complete black hole exterior region, with direct control on which are the sources in the equations of motion. As we will see, this approach will facilitate the microscopic interpretation.

In the last years an intensive effort to understand the effect of higher-curvature corrections to solutions of heterotic string theory has been performed \cite{Cano:2018qev, Chimento:2018kop, Cano:2018brq, Cano:2018hut, Cano:2018aod, Faedo:2019xii, Cano:2019oma, Cano:2019ycn, Ruiperez:2020qda}. The cases considered include different supersymmetric configurations of strings, momentum, Kaluza-Klein monopoles (KK) and solitonic 5-branes (S5), as well as some non-extremal black holes (that lack a microscopic interpretation to date). The small-black-hole system with four non-compact dimensions was studied in \cite{Cano:2018hut}, where it was found that the perturbative curvature corrections leave the field theory solution singular. Additionally, a curvature-corrected solution with a regular horizon and whose Wald entropy coincides with \eqref{eq:smallentropy} was described. It was argued that this field configuration should not be identified microscopically with the DH small black hole, because it contains a KK monopole. Interestingly, this charge does not appear explicitly in the entropy formula, although its value needs to differ from zero in order to have a regular horizon. A crucial ingredient in the construction is the presence of localized solitonic 5-brane sources, with a non-trivial charge profile that asymptotes to zero. The presence of these branes is the ultimate reason for the regularity of the horizon. Two important points to notice are that this solution is already regular in the zeroth-order supergravity description, and that the main effect of higher-curvature corrections is to screen the S5-brane charge. Hence, the system describes a modification of the one studied by Dabholkar and Harvey \cite{Dabholkar:1989jt} obtained adding non-perturbative sources, just like Strominger and Vafa did in \cite{Strominger:1996sh}. Since it reproduced the entropy of the DH states but did not match other properties, the solution was called a ``fake'' small black hole in \cite{Cano:2018hut}. Likewise, the perturbative corrections to the small black ring system have been computed in \cite{Ruiperez:2020qda}, showing that the field configuration remains singular after their inclusion. 

Motivated by these results, we perform here a more exhaustive study of a large family of supersymmetric solutions of the heterotic theory, which includes singular and regular horizon black holes, as well as other related configurations that serve to gain a broader perspective on the matter. We will not restrain ourselves to spacetimes with four non-compact dimensions, but will study the problem for $4 \leq d \leq 9$ (DH states describe strings wrapping one compact direction). 

\subsection{Content of the paper}

In section \ref{sec:pert} we describe our default course of action for the construction of solutions of heterotic string theory at first order in $\alpha'$. Section \ref{sec:BHreview} reviews the supersymmetric black holes with four (five) non-compact dimensions that result from the superposition of the four (three, without KK monopole) types of sources considered in the article. Besides describing the complete solutions in the exterior of the black hole, we use the near-horizon entropy function formalism as an alternative method to obtain some relevant properties of the solutions. The purpose of this is manifold; on one side, it is useful as a consistency check and facilitates the comparison with previous literature, while on the other side it is illustrative to show how some information beyond the near-horizon background needs to be given in order to distinguish between solutions with 3 charges and 4 charges with unit KK monopole. In section \ref{sec:KKS5}, we study the fields that result from general superposition of KK monopoles and S5 branes. Special attention is paid to the merging of monopoles and possible emergence of conical defects. It is described that fractional charge contributions induced by curvature corrections is a generic property of orbifolded spaces, consequence of the fact that the integral of the Bianchi identity is related to the orbifold Euler character of the space, which typically has non-integer value. The curvature corrections to small black holes and rings (string with winding and momentum, static or oscillating) in general number of dimensions are computed in section \ref{sec:SBHs}. Finally, section \ref{sec:fakesmall} describes a very special family of black hole solutions of the kind considered in section \ref{sec:BHreview} with the property that the S5 charge is completely screened, which we call \emph{fake small black holes}. Some of the properties of the resulting field configuration (but, crucially, not all of them) coincide with those of the DH states. Most importantly, while DH (and, hence, small black holes) are $1/2$ BPS states, fake small black holes are $1/4$ BPS. It is shown how a supersymmetric solution with regular horizon is only possible if there are at least 5 compact dimensions, which illustrates that the construction is possible due to the presence of S5 branes wrapping the internal space. Some conclusions are collected in section \ref{sec:conc}. Supplementary technical information is contained in the appendices.

\section{On the perturbative approach to stringy solutions}
\label{sec:pert}

It is with relative frequency that problems need to be approached perturbatively. In some occasions the equations that need to be solved are known, but they are too complicated to be directly treated. In that case, it may happen that those can be expressed as a small modification, in some appropriate sense, of a set of simpler equations, for which analytic solutions can be found. The system is then expressed in terms of a series expansion, possibly with infinite terms, where the \emph{zeroth order} term corresponds to the simpler set of equations. Another common situation, which we will encounter in this work, is that only a perturbative description of the system is known, with the complete non-perturbative formulation inexistent or unknown. A schematic representation of such a problem is

\begin{equation}
\label{eq:genpert}
\sum_{n=0}^{\infty} \alpha^n f_{n,i}\left[ \phi^a, \mathcal{O} (\phi^a) \right] =0 \, .
\end{equation}

\noindent
Here $i$ labels a number of independent equations for the variables $\phi^a$, with $\mathcal{O} (\phi^a)$ collectively representing any possible operator acting on the variables. The expansion is controlled by the presence of the parameter $\alpha$, whose power serves to label the order of the correction. The functional form of terms of higher order $n>k$ could be unknown, or simply it may be computationally convenient to truncate the series at a certain order. Perturbative solutions to the system at $kth$-order are expressions of the form

\begin{equation}
\label{eq:gensol}
\phi^a= \phi^a_0+\sum_{n=1}^{k} \alpha^n \phi^a_n \, ,
\end{equation}

\noindent
such that, when substituted in \eqref{eq:genpert}, the equations are not necessarily identically satisfied, but the non-vanishing terms are of order $k+1$ or higher in the expansion parameter. Such expression is usually interpreted as a good approximation of the real solution of the full system if some requirements are fulfilled, which include an estimation of how \emph{small} the non-vanishing part of the equations is.

The zeroth-order term in \eqref{eq:gensol}, $\phi^a_0$, plays a special role. It is an exact solution of the zeroth-order system of equations that serves as the starting point in the construction of the solution. In order to obtain it, boundary conditions need to be given for the variables. These boundary conditions are considered to be part of the specification of the zeroth-order system. The subsequent terms in the expansion $\phi^a_k$ are progressively computed using the previously obtained values for $\phi^a_{m}$, with $m<k$, as input in equation \eqref{eq:genpert}, which is then solved up to terms of order $\alpha^{k+1}$. The perturbative solution is therefore built order by order from $\phi^a_0$, which can be used as a sort of label to identify the configuration. 

The variables we shall be interested in are fields defined on a manifold. In the problems we find in this article, boundary conditions can be chosen following different approaches. In first place, we need to specify the asymptotic structure of the manifold (i.e. its topology) and the assumed isometries. The remaining information can be specified through the introduction of localized sources in the equations of motion or, alternatively, indicating the asymptotic fall-off behaviour of (independent) fields. In this work, this corresponds either to the election of sources signalling the presence of fundamental objects of the heterotic theory, or the independent charges carried by the field configuration. Both possibilities are technically valid and, most frequently, they define inequivalent perturbative expansions. The reason is that, in certain configurations, some of the higher-order terms behave as delocalized sources of charge in the equations of motion, affecting the original relation between localized sources and asymptotic charges of the fields. Hence, if one of these properties is kept constant in the construction of the perturbative solution, the other one will change, and viceversa. When constructing these solutions, it is fundamental to identify these relations appropriately and understand their implications, as we emphasize at several stages in this work. 

In the perturbative constructions presented below, the boundary conditions are fixed by specifying the localized sources in the system. The advantage of this approach is that, for the systems in which there is a string theory interpretation, the fundamental constituents of the solution remain fixed, so it gives us information of how higher-curvature corrections modify a given stringy configuration.

\subsection{Heterotic theory}
\label{sec:theory}

The low-energy limit of heterotic string theory is described by an effective field theory for its massless modes ---the metric $g_{\mu\nu}$, the dilaton $\phi$, the Kalb-Ramond (KR) 2-form $B_{\mu\nu}$, and a set of non-Abelian Yang-Mills fields $A^{A}_{\mu}$ with gauge group fixed to be either SO(32) or ${\mathrm E}_{8}\times {\mathrm E}_{8}$---\footnote{We will however work with a consistent truncation in which all the Yang-Mills vector fields are trivial.} which involves a double perturbative expansion in $\alpha'$, the string length square, and $g_{s}$, the string coupling. In this work, we will only deal with the $\alpha'$ expansion, assuming we are in a regime where $g_{s}$- or loop corrections can be neglected.\footnote{Of course, this is something that must be checked \emph{a posteriori}. The solutions described in Section \ref{sec:KK+S5} have a divergent dilaton, we refer to \cite{CALLAN1991611} for more information about this issue. }

\subsubsection{Effective action and equations of motion}

To first order in $\alpha'$, the bosonic part of the effective action of the heterotic string is given by \cite{Gross:1986mw, Bergshoeff:1989de, Metsaev:1987zx}

\begin{equation}
\label{action}
{S}
=
\frac{g_{s}^{2}}{16\pi G_{\rm N}^{(10)}}
\int d^{10}x\sqrt{|{g}|}\, 
e^{-2{\phi}}\, 
\left[
{R} 
-4(\partial{\phi})^{2}
+\frac{1}{2\cdot 3!}{H}^{2}
-\frac{\alpha'}{8}R_{(-)}{}_{\mu\nu}{}^a{}_b R_{(-)}{}^{\mu\nu\, b}{}_a +\dots
\right]\, ,
\end{equation}

\noindent 
where  $R_{(-)}{}^a{}_b = d\omega_{(-)}{}^a{}_b - \omega_{(-)}{}^a{}_c \wedge \omega_{(-)}{}^c{}_b$ is the curvature of the torsionful spin connection, defined as $\omega_{(-)}{}^a{}_b\equiv\omega^a{}_b-\frac{1}{2}H_{c}{}^a{}_b\, e^c$, where $\omega^a{}_b$ is the spin connection. The 3-form field strength $H$ associated to the Kalb-Ramond 2-form $B$ is given by 

\begin{equation}
\label{def:kalb-ramond}
H = dB+ \frac{\alpha'}{4} \Omega^{\text{L}}_{(-)} \,,
\end{equation}

\noindent 
where

\begin{equation}
\label{def:chern-simons}
\Omega^{\text{L}}_{(-)} = d\omega_{(-)}{}^a{}_b \wedge \omega_{(-)}{}^b{}_a - \frac{2}{3} \omega_{(-)}{}^a{}_b \wedge \omega_{(-)}{}^b{}_c \wedge \omega_{(-)}{}^c{}_a \,,
\end{equation}

\noindent
is the Chern-Simons 3-form of $\omega_{(-)}{}^a{}_b$. The Bianchi identity is obtained by taking the exterior derivative of eq.~\eqref{def:kalb-ramond}, getting 

\begin{equation}\label{eq:bianchi}
dH=\frac{\alpha'}{4}R_{(-)}{}^a{}_b\wedge R_{(-)}{}^b{}_a \, .
\end{equation}

The equations of motion at first order in $\alpha'$ can be obtained by varying the action \eqref{action} with respect to the metric, dilaton and Kalb-Ramond 2-form. In doing so, we can ignore \emph{implicit} occurrences of these fields through the torsionful spin connection, which according to the Bergshoeff-de Roo lemma yield terms of second order in $\alpha'$ \cite{Bergshoeff:1989de}.\footnote{It is worth to emphasize that this only holds if one works perturbatively in $\alpha'$.}  The set of equations that one obtains is

\begin{eqnarray}
\label{eq:eq1}
R_{\mu\nu} -2\nabla_{\mu}\partial_{\nu}\phi
+\frac{1}{4}{H}_{\mu\rho\sigma}{H}_{\nu}{}^{\rho\sigma}
-\frac{\alpha'}{4}R_{(-)}{}_{\mu\rho}{}^a{}_b R_{(-)}{}_\nu{}^{\rho\, b}{}_a
& = & 
\mathcal{O}(\alpha'^2)\, ,
\\
& & \nonumber \\
\label{eq:eq2}
(\partial \phi)^{2} -\frac{1}{2}\nabla^{2}\phi
-\frac{1}{4\cdot 3!}{H}^{2}
+\frac{\alpha'}{32}R_{(-)}{}_{\mu\nu}{}^a{}_b R_{(-)}{}^{\mu\nu\, b}{}_a
& = &
\mathcal{O}(\alpha'^2)\, ,
\\
& & \nonumber \\
\label{eq:eq3}
d\left(e^{-2\phi}\star\!{H}\right)
& = &
\mathcal{O}(\alpha'^2)\, .
\end{eqnarray}

\noindent
Although it is not explicitly written in \eqref{eq:bianchi}-\eqref{eq:eq3}, the equations are allowed to have localized sources in the form of Dirac delta functions. These appear at zeroth-order in the perturbative expansion of the theory. An election of sources correspond to a choice of boundary conditions, being ultimately responsible for the zeroth-order background studied. The possibility that higher-order corrections induce new terms of this form should not a priori be discarded (that would mean that the first-order terms produce a Dirac delta function when evaluated on the zeroth-order background), although this does not occur in the cases we consider here. Hence, when a solution is described perturbatively, the sources are fixed. A modification of those is interpreted as a non-perturbative modification of the background. 

\subsubsection{Supersymmetry transformations}

Later on, we shall be interested in studying the conditions that must be satisfied by our configurations in order to preserve a certain amount of supersymmetry. Therefore, we need to know the supersymmetry transformations of the fermionic fields, the gravitino $\psi_{\mu}$ and the dilatino $\lambda$. Their explicit form also receive $\alpha'$ corrections, but fortunately to us, they appear at cubic order in $\alpha'$, see for instance \cite{Bergshoeff:1989de}. Hence, for the purposes of this work, the supersymmetry transformations of the fermionic fields reduce to

\begin{eqnarray}
\label{eq:deltapsi}
\delta_{\epsilon}\psi_{\mu} 
&=&
\left(
\partial_{\mu} 
-\frac{1}{4}\omega_{(+)\, \mu ab}\,\Gamma^{ab}
\right)\epsilon\, ,
\\
\label{eq:deltalambda}
\delta_{\epsilon}\lambda&=&\left(\partial_{a}\phi\,\Gamma^a -\frac{1}{12}H_{abc}\,\Gamma^{abc}
\right)\epsilon\, ,
\end{eqnarray}

\noindent
where $\omega_{(+)}{}^a{}_b=\omega^a{}_b+\frac{1}{2}H_{c}{}^a{}_b\, e^c$.

\section{Review of regular supersymmetric black holes}
\label{sec:BHreview}

\subsection{Zeroth-order description}\label{sec:zeroBH}

Regular supersymmetric black-hole solutions to supergravity theories have five or four non-compact dimensions\footnote{Higher-dimensional supersymmetric solutions may describe black strings with a null isometry in a non-compact direction of spacetime.}. The simplest black holes of this kind that one can obtain as solutions of the effective equations of motion of the heterotic string have the following form \cite{Cvetic:1995uj, Cvetic:1996xz}

\begin{eqnarray}
\label{eq:10dmetric0thorder} \nonumber
ds^2&=&\frac{2}{\zm}du\left(dt-\frac{\zp}{2}du\right)-\zz \, d\sigma^2-dz^\alpha dz^\alpha \ ,\\ \nonumber
\label{eq:10dH0thorder}
H&=&\star_\sigma d\zz+d\zm^{-1}\wedge du\wedge dt \ ,\\
\label{eq:10ddilaton0thorder}
e^{2\phi}&=&e^{2\phi_\infty}\frac{\zz}{\zm}\ ,
\end{eqnarray}

\noindent
where 

\begin{equation} \label{eq:GHmetric}
d\sigma^2={\cal H}^{-1}\left(d\eta+\chi\right)^2+{\cal H}\,d{\vec x}^2_{(3)}\, , \hspace{1cm} d\chi=\star_{(3)}d{\cal H}\, ,
\end{equation}

\noindent
is the metric of a four-dimensional Gibbons-Hawking (GH) space \cite{Gibbons:1979zt, Gibbons:1987sp}, where the functions ${\cal Z}_{0,+,-}$ are defined. It turns out that this field configuration, as it stands, preserves at least four of the sixteen supersymmetries of the theory. 

The coordinates $z^{\alpha}\sim z^{\alpha}+2\pi \ell_{s}$ parametrize a four-dimensional torus, ${\mathbb T}^{4}$, with no internal dynamics, whereas $z\equiv t-u~\sim z+2\pi R_{z}$ parametrizes an internal direction, ${\mathbb S}^{1}_{z}$, whose dynamics is non-trivial. 

The equations of motion are satisfied if ${\cal Z}_{0,+,-}$ are harmonic functions in the GH space.\footnote{Note that the Gibbons-Hawking function ${\cal H}$ is also harmonic in ${\mathbb E}^{3}$ and in GH.} The choices that yield the black-hole solutions we are interested in are

\begin{equation}
{\cal Z}_{0,+,-}=1+\frac{q_{0,+,-}}{r}\, , \hspace{1cm} {\cal H}=\epsilon+\frac{q_{ H}}{r}\, ,
\end{equation}

\noindent
where $\epsilon$ is either $0$ or $1$ and where $r$ denotes the radial coordinate of ${\mathbb E}^{3}$: $r^2\equiv{\vec x}_{(3)}\cdot{\vec x}_{(3)}$. The  $\epsilon=0$ and $\epsilon=1$ cases will give rise to five- and four-dimensional black holes respectively. Let us consider each case separately.

\subsubsection*{Static, spherically-symmetric, three-charge black holes in five dimensions}

In the $\epsilon=0$ case, which implies there are five non-compact dimensions, the change of variables $\rho^2=4q_{H}r$ and $\displaystyle\psi=\frac{\eta}{q_{H}}$ allows us to rewrite the metric as

\begin{equation}\label{eq:R4metric}
d\sigma^2=d\rho^2+ \frac{\rho^2}{4}\left(d\psi^2+d\phi^2+d\theta^2+2\cos \theta\, d\psi d\phi\right)\, ,
\end{equation}

\noindent
where one can recognize the factor multiplied by $\rho^2$ as the metric of the round 3-sphere ${\mathbb S}^{3}$. Hence, this (trivial) choice of the GH function gives four-dimensional Euclidean space ${\mathbb E}^{4}$, provided $\psi \in (0,4\pi), \phi \in (0,2\pi)$ and $\theta \in (0,\pi)$. With the solution written in this form, the parameter $q_H$ has disappeared from the field expressions, and one concludes that it has no physical significance in these configurations (it merely sets the scale of a change of coordinates).

After dimensional reduction over the internal directions, the following metric (in the so-called modified Einstein frame) is obtained 

\begin{equation}
ds^2_{(5)}=\left(\zz\zp\zm\right)^{-2/3}dt^2-\left(\zz\zp\zm\right)^{1/3}\left(d\rho^2+\rho^2\,d\Omega^2_{(3)}\right)\, .
\end{equation}

\noindent
It represents an extremal black hole with three electric charges. The horizon is placed at $\rho=0$ and the Bekenstein-Hawking entropy is given by

\begin{equation}
S_{\rm{BH5d}}=\frac{\pi^2}{2 G_{\rm N}^{(5)}}\sqrt{{\tilde q}_{0}{\tilde q}_{+}{\tilde q}_{-}}\, ,
\end{equation}

\noindent
where $\displaystyle{{\tilde q}_{0,+,-}\equiv 4 q_{H} q_{0,+,-}}$. In addition to the metric, the compactification yields three vector fields, with $A_i=-Z_i^{-1} dt$, and two scalars. The gauge-invariant conserved electric charge carried by a vector field inside a co-dimension 2 compact spacelike surface (usually taken to be a 3-sphere) is defined, up to a normalization constant, as the integral over the surface of the variation of the Lagrangian with respect to the $rt$ component of the field strength\footnote{General five-dimensional supergravities contain gauge Chern-Simons terms that give additional contributions, such that magnetic fields become electric sources. The requirement of spherical symmetry implies that 2-forms cannot have magnetic sources, so these terms do not contribute here.}. In the zeroth-order theory, this is the integral of the dual field strength. The evaluation gives $\mathcal{Q}_i \sim \tilde{q}_i$, which is the reason why the poles of the harmonic functions are often referred as ``charges". However, when corrections are incorporated, the previous definition of charge may include additional terms such that the result is not just the pole of the harmonic function. Being of higher-order in derivatives, these terms become subleading in the asymptotic expansion, and we get

\begin{equation}
\lim_{r \rightarrow \infty} \mathcal{Z}_i = 1 + \frac{c_i \mathcal{Q}_i }{r} + \mathcal{O}(r^{-2}) \, ,
\end{equation}

\noindent
for some convenient normalization constants $c_i$, whose value can be inferred from the discussion below. This issue is treated with greater care in the following subsection.

Two of the three types of  charges ---namely, ${\mathcal Q}_{-}$ and ${\mathcal Q}_{0}$--- correspond to the electric and magnetic (or S5-brane) charge associated to the Kalb-Ramond 2-form $B_{\mu\nu}$. They can be understood as being produced by fundamental strings, which are electrically-charged with respect to $B_{\mu\nu}$, and by a stack of $N$ solitonic five-branes, which carry instead magnetic charge. In the lower-dimensional description, these objects act as point-like sources, as they are wrapped along the internal directions. Concretely, fundamental strings are wrapped along ${\mathbb S}^1_{z}$ with total winding number $w$ and solitonic five-branes wrap the five-dimensional torus ${\mathbb T}^{4}\times {\mathbb S}^1_z$. Finally, the charge ${\mathcal Q}_{+}$ is associated to the momentum $n$ of a gravitational wave which travels along the $z$ direction. Introducing these sources in the equations of motion via Dirac delta functions with appropriate coefficients \cite{Cano:2018qev, Cano:2018brq}, it is possible to obtain a relation between the parameters $\tilde q_i$ and the \emph{number} of fundamental objects in the microscopic interpretation,

\begin{equation}\label{eq:stringyid5d}
{\tilde q}_{-}=g^{2}_{s}\alpha' w \, , \hspace{1cm} {\tilde q}_{0}=\alpha' N \, , \hspace{1cm} {\tilde q}_{+}= \frac{g^2_{s}\alpha'^2}{R^2_{z}}n\, .
\end{equation}

\noindent
In this case, the charges of the system are equal to the localized sources, $\mathcal{Q}_{-}=w$, $\mathcal{Q}_{0}=N$,  $\mathcal{Q}_{+}=n$. However, it is important to bear in mind that, in general, these quantities are different, as will become evident when we include $\alpha'$ corrections.  
Now, using

\begin{equation}
G_{\rm N}^{(5)}=\frac{G_{\rm N}^{(10)}}{(2\pi \ell_{s})^4 2\pi R_{z}}=\frac{\pi g_{s}^2 {\alpha'}^2}{4R_{z}}\, , \hspace{1cm} (G^{(10)}_{\rm N}=8\pi^6 g_{s}^2{\alpha'}^4)\, ,
\end{equation}

\noindent
the Bekenstein-Hawking entropy gives

\begin{equation}
S_{\rm {BH5d}}=2\pi\sqrt{nwN}=2\pi\sqrt{{\mathcal Q}_{+} {\mathcal Q}_{-} {\mathcal Q}_{0}}\, ,
\end{equation}

\subsubsection*{Static, spherically-symmetric, four-charge black holes in four dimensions}

Let us now consider the $\epsilon=1$ case, which means there are four non-compact dimensions. The Gibbons-Hawking 1-form $\chi$ is determined by solving eq.~\eqref{eq:GHmetric}. A possible local expression is

\begin{equation}
\chi=q_{H} \cos \theta \,d\phi\, ,
\end{equation}

\noindent
where we have introduced the spherical coordinates $\theta$ and $\phi$, defined in terms of the Cartesian coordinates ${\vec x}_{(3)}=(x^1, x^2, x^3)$ as

\begin{equation}
x^1=r\sin\theta \cos \phi\, ,\hspace{1cm} x^2=r\sin\theta \sin \phi\, , \hspace{1cm} x^3=r\cos\theta\, .
\end{equation}

\noindent
The resulting metric has a Dirac-Misner string singularity, as $\chi$ is ill-defined at $\theta=0,\pi$. It is well-known that this string can be removed if $\eta$ is a compact coordinate ($\eta\sim \eta+2\pi R_{\eta}$) and $q_{H}$ obeys the quantization condition 

\begin{equation}
q_{H}=\frac{R_{\eta} W}{2}\, , \hspace{1cm} W\in {\mathbb N}\, .
\end{equation}

\noindent
The resulting GH metric describes an orbifold with a conical singularity for integer values of $W$ other than $1$. This can be seen by studying the $r\to 0$ limit of \eqref{eq:GHmetric}, which is

\begin{equation} \label{eq:KKcenter}
d\sigma^2\Big|_{r\to0}=d\rho^2+\frac{\rho^2}{4}\left[\left(d\left(\Psi/W\right)+\cos\theta d\phi\right)^2+d\theta^2+\sin^2\theta d\phi^2\right]\, , 
\end{equation}

\noindent
where we have introduced the radial coordinate $\rho^2=2 R_{\eta} W r$ and the angular coordinate $\Psi=2\eta/R_{\eta}$, whose periodicity is $4\pi$. Then, we see that near $r=0$ the GH metric is that of  the orbifold ${\mathbb E}^{4}/{\mathbb Z}_{W}$. This conical singularity, however, is not present in the ten-dimensional metric \eqref{eq:10dmetric0thorder} as long as ${\tilde q}_{0}\neq0$, in which case the conformal factor behaves as ${\cal Z}_{0}|_{r\to0}\sim {\tilde q}_{0}/\rho^2$ and we are left with the metric of the lens space ${\mathbb S}^{3}/{\mathbb Z}_{W}$, which is perfectly regular. 

The four-dimensional geometry (in modified Einstein frame) that one obtains when compactifying the solution over the internal manifold ${\mathbb T}^{4}\times {\mathbb S}^1_z\times {\mathbb S}^1_\eta$ is the following 

\begin{equation}
ds^2_{(4)}=\left(\zz\zp\zm {\cal H}\right)^{-1/2}dt^2-\left(\zz\zp\zm{\cal H}\right)^{1/3}\left(dr^2+r^2\,d\Omega^2_{(2)}\right)\, ,
\end{equation}

\noindent
and describes an extremal black hole with four charges: $\mathcal{Q}_{0}$, $\mathcal{Q}_{+}$, $\mathcal{Q}_{-}$ and $\mathcal{Q}_{H}=W$. The additional charge with respect to the five-dimensional case, $\mathcal{Q}_{H}$, is the magnetic charge of the Kaluza-Klein vector associated to the compactification over the isometric direction of the GH space. Therefore, the system described by this black-hole solution contains, apart from extended objects present in the five-dimensional case, a KK monopole. The relation between the parameters $q_i$ and the number of fundamental objects is now given by \footnote{The reason for the modification with respect to the expressions in the five-dimensional solution is that the stringy objects are now \emph{smeared} over the transverse direction $\eta$ which forms part of the GH space.}

\begin{equation}\label{eq:stringyidentification}
{q}_{-}=\frac{g^{2}_{s}\alpha'w}{2 R_{\eta}}  \, , \hspace{1cm} {q}_{0}=\frac{\alpha' N}{2R_{\eta}} \, , \hspace{1cm} {q}_{+}= \frac{g^2_{s}\alpha'^2 n}{2R^2_{z} R_{\eta}}\, , \hspace{1cm} q_{H}=\frac{R_{\eta} W}{2}\, .
\end{equation}

\noindent
The horizon of these black holes is again at $r=0$ and the Bekenstein-Hawking entropy in terms of the source parameters and the charges reads

\begin{equation}\label{eq:entropy4d0thorder}
S_{\rm {BH4d}}=2\pi \sqrt{nwNW}=2\pi\sqrt{{\mathcal Q}_{+} {\mathcal Q}_{-} {\mathcal Q}_{0}{\mathcal Q}_{H}} \, ,
\end{equation}

\noindent 
where we have made use of eq.~\eqref{eq:stringyidentification}.


\subsection{First-order description. S5-brane charge screening.} \label{sec:firstBH}

The first-order $\alpha'$ corrections to the black holes presented in the previous section have been computed in an analytic fashion in \cite{Cano:2018qev, Chimento:2018kop, Cano:2018brq}. Let us summarize here the main results of these papers.

A first result is that the corrected solutions have the same form as the zeroth-order expressions given in eqs.~\eqref{eq:10ddilaton0thorder}-\eqref{eq:GHmetric}. This fact can be interpreted as a consequence of supersymmetry, which strongly constrains the form of the field configuration. The functions $\zp$ and $\zz$ receive $\alpha'$ corrections, while $\zm$ and ${\cal H}$ remain unmodified.  In the five-dimensional case, the corrections to these functions take the following form 

\begin{eqnarray} \nonumber
\mathcal Z_+&=&1+\frac{\tilde q_+}{\rho^2}+ \alpha'\frac
{2{\tilde q}_+ \left(\rho^2+{\tilde q}_0+{\tilde q}_-\right)}{{\tilde q}_0 \left(\rho^2+{\tilde q}_0\right)\left(\rho^2+{\tilde q}_-\right)}+\mathcal O(\alpha'^2) \ ,\\
\label{eq5:z0corr5dbhs}
\mathcal Z_0&=&1+ \frac{{\tilde q}_0}{\rho^2}-\alpha' \frac{\rho^2+2{\tilde q}_0}{(\rho^2+{\tilde q}_0)^2}+\mathcal O(\alpha'^2)\,,
\end{eqnarray}

\noindent
whereas in the four-dimensional case it was found that they are given by

\begin{eqnarray} \nonumber
\label{eq:exp_Z+2}
\mathcal{Z}_{+} 
& = &
1+\frac{q_{+}}{r}+ \frac{\alpha' q_{+}}{2q_{H}q_{0}} 
\frac{r^{2}+r\left(q_{0}+q_{-}+q_{H}\right)+q_{H} q_{0}+q_{H} q_{-}+q_{0}q_{-}
}{\left(r+q_{H}\right)\left(r+q_{0}\right)\left(r+q_{-}\right)} 
+\mathcal{O}(\alpha'{}^{2})\, ,
\\
\label{eq:exp_Z02}
\mathcal{Z}_{0} 
& = &
1+\frac{q_{0}}{r}- \alpha'\left[F(r;q_{0})+F(r;q_{H})\right]+\mathcal{O}(\alpha'{}^{2}) \, ,
\end{eqnarray}

\noindent
where 

\begin{equation}
 F(r;q)\equiv\frac{\left(r+q_{H}\right)\left(r+2q\right)+q^{2}}{4q_{H}\left(r+q_{H}\right)\left(r+q\right)^{2}}\, .
\end{equation}

\noindent
On the one hand, the new terms in the functions $\zz$ and $\zp$ are everywhere finite, which implies that the $\alpha'$ corrections do not change neither the location of the horizon nor the near-horizon geometry. This is consequence of the fact that $R_{(-)}{}^a{}_b$ vanishes in these limits, so at this order the equations of motion remain uncorrected in that region. On the other hand, as we describe below, the value of the charges is modified.

Before doing so, it is worth noticing a subtle point about the perturbative construction of the solution. As we constructed it, the near-horizon geometry is entirely determined by the choice of boundary conditions, in terms of delta functions, when solving the equations of motion. This is convenient in the case at hand, as these functions have a direct interpretation in the microscopic theory in terms of localized sources of fundamental objects. Still, it would be possible to follow an alternative approach and solve the equations of motion fixing the boundary conditions in terms of asymptotic properties of the solution. With that choice, one would keep the charges fixed, but the near-horizon geometry and the localized sources would be modified. Since these sources are in correspondence with the microscopic interpretation of the solution, this alternative approach is inconvenient if one wants to study how a given string theory system behaves when $\alpha'$ corrections are incorporated\footnote{A clear example of the importance of this observation can be found in the study of isolated KK monopole solution, which dates back to the $90'$s \cite{Sen:1997zb}, as emphasized in section \ref{sec:KK}. This remark plays a central role in our discussion of small black holes.}. This is the reason why we fix the boundary conditions at $r=0$.

We begin now the discussion about the charges that receive corrections. First, we start noting that in presence of higher-curvature terms the definition is not unique. Consider the Bianchi identity \eqref{eq:bianchi}. In presence of external sources, it is modified as 

\begin{equation}\label{eq:bianchiwithsources}
dH-\frac{\alpha'}{4}R_{(-)}{}^a{}_b\wedge R_{(-)}{}^b{}_a=\star J_{S5} \, ,
\end{equation}

\noindent
where $J_{S5}$ is a six-form current satisfying the conservation law $d\star J_{S5}=0$, which follows from the well-known fact that $R_{(-)}{}^a{}_b\wedge R_{(-)}{}^b{}_a=d\Omega^{\rm L}_{(-)}$. In the case at hand, this current is produced by a stack of S5-branes and its integral over the transverse space to the S5-branes, ${\mathcal M}$, is proportional to the number of S5-branes, $N$:

\begin{equation}
\int_{\mathcal M} \left(dH-\frac{\alpha'}{4}R_{(-)}{}^a{}_b\wedge R_{(-)}{}^b{}_a\right)=\int_{\mathcal M}\star J_{S5}=4\pi^2\alpha' N\, .
\end{equation}

\noindent
Therefore, the relation between the harmonic poles (${\tilde q}_{0}$ and $q_{0}$) and $N$ can be obtained after evaluation of the left-hand side of this equation. As the sources are fixed, one obtains the same result than in previous section \cite{Cano:2018qev, Cano:2018brq},

\begin{equation}\label{eq:Q0=N}
{\tilde q}_{0}=\alpha' N\, , \hspace{1cm} q_{0}=\frac{\alpha' N}{2 R_{\eta}}\, .
\end{equation}

\noindent
The number of S5-branes $N$ coincides with the notion of brane-source charge of ref.~\cite{Marolf:2000cb}. It is worth stressing that the notion of brane-source charge that we have just defined is conserved.  This is different from what happens in other scenarios, like in type II, where the Chern-Simons terms appearing in the Bianchi identities of the RR field strengths are not necessarily closed forms if one allows for external sources. 

In addition, one can define a second notion of charge, which is usually called the Maxwell charge. Unless otherwise stated, this is the notion that we use in the article when we talk about charges --- as opposed to \emph{sources}. In the S5-brane case, it is denoted by ${\cal Q}_{0}$ and it is given by 

\begin{equation}\label{eq:MaxwellS5charge}
{\cal Q}_{0}=\frac{1}{4\pi^2\alpha'} \int_{\partial {\cal M}}H\, .
\end{equation}

\noindent
As the brane-source charge, the Maxwell charge is also conserved and gauge-invariant. The main difference between these two charges lies in the fact that the brane-source charge is localized while the Maxwell charge is not, as it gets contributions from the quadratic-curvature corrections, which behave as effective delocalized sources of S5-brane charge in the Bianchi identity. Evaluating \eqref{eq:MaxwellS5charge} and using \eqref{eq:Q0=N}, in the five-dimensional case we obtain

\begin{equation} \label{eq:S5charge5}
{\cal Q}_{0}=\frac{{\tilde q}_{0}-\alpha'}{\alpha'}=N-1\, ,
\end{equation}

\noindent
while in the four-dimensional case,

\begin{equation} \label{eq:S5charge4}
{\cal Q}_{0}=\frac{2 R_{\eta}}{\alpha'}\left(q_{0}-\frac{\alpha'}{2q_{H}}\right)=N-\frac{2}{W}\, .
\end{equation}

\noindent
We observe that, up to a normalization constant, the Maxwell charge can be read from the asymptotic expansion of the function $\mathcal{Z}_0$, with $\mathcal{Z}_0 = 1+c_0 \mathcal{Q}_0 /r + \mathcal{O}(r^{-2})$ for large $r$. Hence, we see that this also gives the charge carried by the lower-dimensional vectors, as defined in previous subsection. 

Let us now discuss the momentum charge. Analogously to the S5-brane case, the coefficients ${\tilde q}_{+}$ and $q_{+}$ ---which control the leading (divergent) term of the function ${\cal Z}_{+}$ in the near-horizon limit--- are related to the momentum $n$ exactly as in the zeroth-order solution, see eqs.~\eqref{eq:stringyid5d} and \eqref{eq:stringyidentification}. The $\alpha'$ corrections to this function give a contribution to the asymptotic charge carried by the lower-dimensional Kaluza-Klein vector, which can be read off from the asympotic expansion of ${\cal Z}_{+}$, which is ${\cal Z}_{+}=1+c_{+}{\mathcal Q}_{+}/r+ {\cal O}(r^{-2})$, with: 

\begin{equation}
{\cal Q}_{+}=n\left(1+\frac{2}{N}\right)\, ,
\end{equation}

\noindent
in the five-dimensional case and 

\begin{equation}
{\cal Q}_{+}=n\left(1+\frac{2}{NW}\right)\, ,
\end{equation}

\noindent
in the four-dimensional case. An interesting task for the future would be to investigate how these two different notions of Kaluza-Klein momentum charge appear when compactifying the $\alpha'$-corrected action \eqref{action} on a circle \cite{Elgood:2020xwu, Ortin:2020xdm}. The remaining charges do not receive corrections, hence
\begin{equation}
\mathcal{Q}_{-}=w \, , \qquad \qquad \mathcal{Q}_{H}=W\, ,
\end{equation}

\noindent
as in the zeroth-order solution.

It is possible to compute the entropy of these black holes using directly Wald's formula. The presence of the Riemann curvature tensor in the field strength $H$ makes this a subtle problem, which has been addressed in previous literature\footnote{In general, the application of Wald's formula to the heterotic theory gives a gauge dependent expression. This problem can be solved for the family of solutions considered, where is possible to write the action in a covariant form after imposing symmetries and adding boundary terms. In order to deal with Chern-Simons terms, the approach takes the dual Kalb-Ramond field strength (which transforms as a tensor) as the fundamental field, see \cite{Faedo:2019xii}.} \cite{Sen:2005wa, Sahoo:2006pm}. For the family of black holes we are interested in, the result was obtained in \cite{Faedo:2019xii}. For the five- and four-dimensional solutions it was found, respectively

\begin{eqnarray}
\label{eq:waldent5}
S_{\rm {W} 5d} &=& 2\pi \sqrt{nwN}\left(1+\frac{2}{N} \right)=2\pi\sqrt{{\mathcal Q}_{+} {\mathcal Q}_{-} \left({\mathcal Q}_{0}+3 \right) } \, , \\ \label{eq:waldent4}
S_{\rm {W} 4d} &=& 2\pi \sqrt{nwNW}\left(1+\frac{2}{NW} \right)=2\pi\sqrt{{\mathcal Q}_{+} {\mathcal Q}_{-} \left({\mathcal Q}_{0}{\mathcal Q}_{H}+4 \right) }  \, .
\end{eqnarray}

\noindent
The expressions in terms of the sources coincide for both kind of solutions when $W=1$, which is consequence of the fact that their near-horizon limit is identical, i.e. $AdS_3 \times \mathbb{S}^3 \times \mathbb{T}^4$. In other words, it is not possible to distinguish a $4d$ black hole with $\mathcal{Q}_H=W=1$ from a $5d$ black hole if only the near-horizon fields are obtained. Indeed, the distinction is only possible if information beyond the near-horizon region is somehow taken into account.\footnote{The configuration given in \eqref{eq:nH4c} with $W=1$ can be interpreted in two different manners: a near-horizon background of a black hole with three or four independent charges. However, each interpretation is only consistent when, outside the horizon, there is respectively a five- or four-dimensional non-compact space. } In the presence of a general KK monopole of charge $W$, one can write in convenient coordinates

\begin{eqnarray}
\nonumber
ds^{2}
& = &
\frac{\rho^2}{g_s^2 \alpha' w}du
\left[dt-\frac{g_s^2 \alpha' n}{R_z^2 \rho^2} du\right]-
\alpha' N W \left[ \frac{d\rho^2}{\rho^2}+d\Omega^2_{(3)/\mathbb{Z}_W} \right]
-d\vec{z}^2_{(4)}\, ,
\\
& & \nonumber \\ \nonumber
e^{-2{\phi}}
& = &
\frac{w}{N}\, ,
\\
& & \nonumber \\
\label{eq:nH4c}
H 
& = & 
\frac{1}{g_s^2 \alpha' w} \rho d\rho \wedge du \wedge dt+\frac{\alpha' N}{4} \sin\theta d\theta  \wedge d\psi \wedge d\varphi
\, .
\end{eqnarray}

 Heterotic string theory on this background was studied in \cite{Kutasov:1998zh}, where the left and right central charges were determined to be $c_l=6 \mathcal{Q}_{-} (k+2)$ , $c_r= 6 \mathcal{Q}_{-} k$, with $k$ the total $\widehat{SL(2)}$ level for the right-movers ($k+2$ for the left-movers). Upon use of Cardy's formula, the microscopic entropy  (to all orders in the $\alpha'$ expansion in the large charge approximation) that one obtains is
\begin{equation}
\label{eq:cardyent}
S_{\rm {C}} = 2\pi \sqrt{ {\mathcal Q}_{+} {\mathcal Q}_{-} (k +2) } \, .
\end{equation}

\noindent
We notice that this expression matches both \eqref{eq:waldent5} and \eqref{eq:waldent4} if the level is identified with the $AdS$ curvature radius in string units as $k=NW$ (with the understanding that $W=1$ in the five-dimensional case). As observed in \cite{Kutasov:1998zh}, consistency of the bosonic $\widehat{SU(2)}$ CFT theory on this background requires that its level, which was found to be $k-2$, must be the product of $W$ and another integer, which implies that ${\mathcal Q}_{0}$ is quantized. Our expressions \eqref{eq:waldent4} and \eqref{eq:waldent5} in terms of the charges also match those of \cite{Sen:2007qy} (identically) and the perturbative expansion obtained in \cite{Castro:2008ys}, respectively. Since $R_{(-)}\,^a\,_b$ is zero in the near-horizon background, no corrections are expected in this region in the higher-curvature expansion, such that the expressions \eqref{eq:waldent5} and \eqref{eq:waldent4} in terms of the sources would also be exact in the $\alpha'$ expansion. 

Before continuing, we recall that the derivation of Wald's formula from the first law of black hole mechanics assumes that all fields in the theory behave as tensors under general coordinate transformations, although this is only true for the metric and the dilaton (not for the Kalb-Ramond $B_{\mu\nu}$, which includes gauge and Nicolai-Townsend \cite{NICOLAI1981257} transformations). A proof of the first law taking into account this property of the heterotic theory at first order in $\alpha'$ has been only recently found \cite{Elgood:2020nls, Elgood:2020mdx}, obtaining a manifestly gauge invariant general entropy formula. When applied to the solutions at hand, the result reproduces \eqref{eq:waldent5}, \eqref{eq:waldent4}.

In the following section we derive these expressions for the entropies and the charges using a near-horizon approach.

\subsection{Near-horizon entropy function formalism}\label{sec:eff}

Some aspects of the $\alpha'$-corrected black holes we have just presented have been previously studied in \cite{Sen:2005iz, Sahoo:2006pm, Prester:2008iu, Prester:2010cw}  making use of the entropy function formalism developed by Sen et al. \cite{Sen:2005iz, Sen:2007qy}, which provides a useful method to find the near-horizon geometry and the entropy of extremal black holes. The aim of this subsection is to review this formalism and check its compatibility with the results presented in previous subsections. From a ten-dimensional perspective, the near-horizon geometry of both types of black holes is essentially the same: $\rm{AdS}_{3}\times {\mathbb S}_{3}\times {\mathbb T}^4$ in the five-dimensional case and $\rm{AdS}_{3}\times {\mathbb S}_{3}/{\mathbb Z}_{W}\times {\mathbb T}^4$ in the four-dimensional case. Therefore, for most of the discussion it is enough to study the near-horizon geometry of the four-dimensional black holes. The five-dimensional one will be carefully recovered setting $W=1$ and taking into account the implications of having one additional non-compact coordinate in the asymptotic space, such that one obtains \eqref{eq:S5charge5} for the S5-brane charge instead of \eqref{eq:S5charge4}. 


\subsubsection{Leading-order computation}

As a warm-up exercise, let us first consider the leading-order computation, ignoring for the time being the $\alpha'$ corrections. In this approximation, the effective action is that of ten-dimensional ${\cal N}=1$ supergravity compactified on ${\mathbb T}^{4}\times {\mathbb S}^{1}_{z}\times{\mathbb S}^{1}_{\eta}$. The compactification on the trivial ${\mathbb T}^{4}$ is straightforward and yields

\begin{equation}
S=\frac{g_{s}^{2}}{16\pi G_{\rm N}^{(6)}}\int d^6{\hat x}\sqrt{|\hat g|}\, e^{-2\phi}\left(\hat R-4(\partial \phi)^2+\frac{1}{12}H^2\right)\, ,
\end{equation}

\noindent
where $\hat R$ is the Ricci scalar of ${\hat g}_{\hat \mu \hat \nu}$, the six-dimensional metric, and ${\hat x}^{\hat \mu}=\left\{x^{\mu}, z, \eta\right\}$, with $\mu=\{0, 1, 2, 3\}$, denote the coordinates of the six-dimensional spacetime. Further compactification on $z$ and $\eta$ yields the STU model of ${\cal N}=2, d=4$ supergravity

\begin{equation}
S=\frac{g_{s}^{2}}{16\pi G_{\rm N}^{(4)}}\int d^4x\sqrt{|g|}\,s\left(R-a_{ij}\partial_{\mu}\phi^i \partial^\mu \phi^j-t^2{F^{(1)}}^2-u^2{F^{(2)}}^2-\frac{u^2}{s^2}{F^{(3)}}^2-\frac{t^2}{s^2}{F^{(4)}}^2\right)\, ,
\end{equation}

\noindent
where $\phi^{i}=\{s, t, u\}$ are the three scalar fields present in this model. The relation between the lower- and the higher-dimensional fields is 

\begin{eqnarray}
\label{eq:dimredmetric} \nonumber
g_{\mu\nu}&=&{\hat g}_{\mu\nu}-\frac{{\hat g}_{\mu z}{\hat g}_{\nu z}}{{\hat g}_{zz} }-\frac{{\hat g}_{\mu \eta}{\hat g}_{\nu \eta}}{{
\hat g}_{\eta \eta}}\, , \\ \nonumber
\nonumber\\ \nonumber
\label{eq:dimredvectors}
A^{(1)}&=&-\frac{{\hat g}_{\mu z}}{2{\hat g}_{zz} }\,, \hspace{0.5cm} A^{(2)}=-\frac{{\hat g}_{\mu \eta}}{2{\hat g}_{\eta\eta} }\, , \hspace{0.5cm} A^{(3)}=\frac{{\tilde B}_{z\mu}}{2}\,, \hspace{0.5cm} A^{(4)}=\frac{{\tilde B}_{\eta\mu}}{2}\, , \\
\nonumber\\
\label{eq:dimredscalars}
s&=&e^{-2\phi}\sqrt{{\hat g}_{zz} {\hat g}_{\eta\eta}}\, , \hspace{0.5cm} t=\sqrt{|{\hat g}_{zz}|}\, , \hspace{0.5cm} u=\sqrt{|{\hat g}_{\eta\eta}|}\, ,
\end{eqnarray}

\noindent
where ${\tilde B}_{\hat \mu \hat \nu}$ is the dual of the Kalb-Ramond 2-form $B_{\hat \mu \hat \nu}$, defined as

\begin{equation}\label{deftildeH}
d{\tilde B}={\tilde H}\equiv e^{-2\phi}\star H\, .
\end{equation}

\emph{\underline{Ansatz for the near-horizon geometry}}. We shall restrict ourselves to study the near-horizon geometry of static, extremal, spherically-symmetric black holes assuming that not only the metric but all fields are invariant under the SO$(2,1)$ $\times$ SO$(3)$ isometry group. The most general ansatz consistent with this symmetry and the four type of charges we want to describe is

\begin{equation}
\begin{aligned}
\label{ansatz4d}
ds^2=&\, v_{1}\left(r^2dt^2-\frac{dr^2}{r^2}\right)-v_{2}\left(d\theta^2+\sin^2\theta d\phi^2\right)\, ,\\
F^{(1)}=&\,e_{1}\, dr\wedge dt\, , \hspace{0.2cm} F^{(2)}=p_{2}\,\sin\theta d\theta \wedge d\phi \, , \hspace{0.2cm} F^{(3)}=e_{3}\, dr\wedge dt\, ,\hspace{0.2cm} F^{(4)}=p_{4}\,\sin\theta d\theta \wedge d\phi \, ,\\
s=&u_{s}\,, \hspace{1cm} t=u_{t}\, , \hspace{1cm} u=u_{u}\, , 
\end{aligned}
\end{equation}

\noindent 
where $v_{1}, v_{2}, e_{1},e_{3}, p_{2}, p_{4}$ and $\vec u\equiv (u_{s},u_{t}, u_{u})$ are constants. The election of electric or magnetic character of $F^{(i)}$ is motivated by the stringy interpretation of the solution.

The above configuration can be straightforwardly uplifted to six dimensions by making use of eqs. \eqref{eq:dimredmetric}. We obtain:

\begin{equation}\label{ansatz6d}
\begin{aligned}
d{\hat s}^2=\, &v_{1}\left(r^2dt^2-\frac{dr^2}{r^2}\right)-v_{2}\left(d\theta^2+\sin^2\theta d\phi^2\right)-u^{2}_{t}\left(dz-2e_{1}r dt\right)^2\\
&-u^{2}_{u}\left(d\eta+2p_{2}\cos\theta d\phi\right)^2\, ,\\
{\tilde H}=\,&2e_{3}\,dt\wedge dr\wedge dz+2p_{4}\,\sin\theta d\theta \wedge d\eta\wedge d\phi\, ,\\
e^{2\phi}=\, &\frac{u_{t}u_{u}}{u_{s}}\, .
\end{aligned}
\end{equation}

\emph{\underline{Extremization of the entropy function}}. Following \cite{Sen:2005wa}, we define the function  $f(v_{1}, v_{2}, {\vec u}, e_{i}, p_{i})$ as the integral over the angular coordinates of the (four-dimensional) Lagrangian evaluated on the ansatz \eqref{ansatz4d}: 

\begin{equation}
f(v_{1}, v_{2}, {\vec u}, e_{i}, p_{i})\equiv \int d\theta d\phi \,(\sqrt{|g|}{\cal L})|_{\eqref{ansatz4d}}\, .
\end{equation}

It can be shown that the metric and scalar equations of motion reduce to the extremization of the function $f$ with respect to $v_{1}, v_{2}$ and $\vec u$, while the equations of motion of the vector fields and the Bianchi identities are trivially satisfied for this ansatz. Hence,  the extremization of the function $f$ gives five equations which fix $v_{1}, v_{2}$ and $\vec u$ in terms of the electric and magnetic charges of the black hole. The latter are defined as

\begin{equation}
Q_{I}=\frac{G_{\rm N}^{(4)}}{g_{s}^2}\int d \theta d\phi \, \frac{\delta}{\delta F^{(I)}_{rt}}(\sqrt{|g|}{\cal L})\, ,\hspace{1cm} P_{I}=\frac{1}{4\pi} \int d\theta d\phi \, F^{(I)}_{\theta \phi}\, .
\end{equation}

\noindent
Then,

\begin{equation}\label{charges}
Q_{I}= \frac{G_{\rm N}^{(4)}}{g_{s}^2}\frac{\partial f}{\partial e_{I}}\, , \hspace{1cm} \text{and}\hspace{1cm} P_{I}=p_{I}\, .
\end{equation}

\noindent
Let us now define the entropy function $\cal E$ as the Legendre transform of $f$,

\begin{equation}
{\cal E}(v_{1},v_{2},\vec u, Q_{I}, P_{I})=2\pi\left(\frac{g_{s}^2}{G^{(4)}_{\rm N}}Q_{I}e_{I}-f(v_{1},v_{2},\vec u, e_{I}, P_{I})\right)\, ,
\end{equation}

\noindent
where the parameters $e_{I}$ should be regarded as functions of the electric charges, $e=e\left(Q\right)$. The entropy function evaluated at the extremum of $f$ gives Bekenstein-Hawking entropy as a function of the electric and magnetic charges \cite{Sen:2005wa},

\begin{equation}
S_{\rm BH}(Q, P)={\cal E}(v_{1, \rm {ext}}(Q, P), v_{2, \rm{ext}}(Q, P), {\vec u}_{\rm{ext}}(Q, P), Q, P)\, .
\end{equation}

In the case at hand, the function $f$ is found to be equal to

\begin{equation}\label{eqfleadingorder}
f\left (v_{1},v_{2}, \vec u, e_{i}, P_{i}\right)=\frac{g^2_{s}}{2 G_{\rm N}^{(4)}} \left[{u_s} \left(\frac{e_1^2 u_t^2 v_2}{{v_1}}-\frac{p_2^2 u_u^2 v_1}{{v_2}}+{v_1}-{v_2}\right)+\frac{e_3^2 u_u^2 v_2^2-p_4^2 u_t^2 v_1^2}{u_s v_{1} v_{2}}\right]\, ,
\end{equation}

\noindent
and it has an extremum at 

\begin{equation}\label{extremum}
\begin{aligned}
v_{1, \rm{ext}}=&v_{2,\rm{ext}}=4Q_{3}P_{2}\, , \hspace{0.5cm}  e_{1, \rm{ext}}=\sqrt{\frac{Q_{3}P_{2}P_{4}}{Q_{1}}}\,, \hspace{0.5cm} e_{3, \rm{ext}}=\sqrt{\frac{Q_{1}P_{2}P_{4}}{Q_{3}}}\, ,\\
\\
{\vec u}_{\rm {ext}}=&\left(\sqrt{\frac{Q_{1}P_{4}}{Q_{3}P_{2}}}, \sqrt{\frac{Q_{1}}{P_{4}}}, \sqrt{\frac{Q_{3}}{ P_{2}}}\right)\, .
\end{aligned}
\end{equation}

\noindent
Substituting the values of $v_{1}, v_{2}$ and $\vec u$ at the extremum of the entropy function in the six-dimensional ansatz \eqref{ansatz6d} yields 
\begin{equation}\label{ansatz6dext}
\begin{aligned}
d{\hat s}^2=\, &4 Q_{3} P_{2}\left(r^2dt^2-\frac{dr^2}{r^2}-d\theta^2-\sin^2\theta d\phi^2\right)- \frac{Q_{1}}{P_{4}}\left(dz-2\sqrt{\frac{Q_{3} P_{2} P_{4}}{Q_{1}}}r dt\right)^2\\
&- \frac{Q_{1}}{P_{4}}\left(d\eta+2 P_{2}\cos\theta d\phi\right)^2\, ,\\
{\tilde H}=\,&2\sqrt{\frac{Q_1 P_2 P_4}{Q_3}}\,dt\wedge dr\wedge dz+2P_{4}\,\sin\theta d\theta \wedge d\eta\wedge d\phi\, ,\\
e^{2\phi}=\, &\frac{Q_3}{P_4}\, .
\end{aligned}
\end{equation}

\noindent
We can now make a comparison with the near-horizon limit of the solutions studied in the previous subsections to extract the relation between the electric and magnetic charges $(Q, P)$ and the source parameters:

\begin{equation}\label{eq:sources}
Q_{1}=\frac{\alpha'^2 n}{4 R^2_{z} R_{\eta}} \, , \hspace{1cm} Q_{3}=\frac{\alpha' N}{4 R_{\eta}}\, , \hspace{1cm} P_{2}=\frac{W R_{\eta}}{4}\, , \hspace{1cm} P_{4}=\frac{\alpha' w}{4 R_{\eta}}\, .
\end{equation}

\noindent
Plugging these values back into \eqref{ansatz6dext}, 

\begin{equation}\label{ansatz6dext2}
\begin{aligned}
d{\hat s}^2=\, &\frac{\alpha' NW}{4}\left(r^2dt^2-\frac{dr^2}{r^2}-d\theta^2-\sin^2\theta d\phi^2\right)- \frac{\alpha'n}{R^2_{z}w}\left(dz-\frac{R_{z}}{2}\sqrt{\frac{w NW}{n}}r dt\right)^2\\
&- \frac{\alpha' N}{R^2_{\eta}W}\left(d\eta+\frac{WR_{\eta}}{2}\cos\theta d\phi\right)^2\, ,\\
{\tilde H}=\,&\frac{\alpha'}{2R_{z}}\sqrt{\frac{nwW}{N}}\,dt\wedge dr\wedge dz+\frac{\alpha' w}{2 R_{\eta}}\,\sin\theta d\theta \wedge d\eta\wedge d\phi\, ,\\
e^{2\phi}=\, &\frac{N}{w}\, ,
\end{aligned}
\end{equation}

\noindent
that matches \eqref{eq:nH4c} after a coordinate redefinition. Let us note an interesting property of the near-horizon limit, which is that the 3-form field strength is selfdual (with respect to the orientation $\epsilon^{tr\theta\phi\eta z}=+1$) in six dimensions, namely $\tilde H=\star \tilde H=e^{-2\phi} H$. This, as we will discuss in section~\ref{sec:fakesmall}, is directly related to supersymmetry.

Finally, the entropy is obtained by evaluating $\cal E$ at the extremum. The function $f$ vanishes there, and we simply have
\begin{equation}
S_{\rm {BH}}(Q_{1},Q_{3}, P_{2}, P_{4})=\frac{2\pi g_{s}^2}{G^{(4)}_{\rm N}}\left(Q_{1}e_{1}+Q_{3}e_{3}\right)|_{\text{ext}}=\frac{4\pi g_{s}^2}{G^{(4)}_{\rm N}}\sqrt{Q_{1}Q_{3} P_{2}P_{4}}=2\pi \sqrt{nw NW}\, ,
\end{equation}

\noindent
in agreement with eq.~\eqref{eq:entropy4d0thorder}.

\subsubsection{First-order $\alpha'$ corrections}

\emph{\underline{Rewriting of the $\alpha'$-corrected action}}. Let us now take into account the $\alpha'$ corrections to the supergravity action \eqref{action}. Since the trivial ${\mathbb T}^{4}$ plays absolutely no role in the discussion, we can directly work in six dimensions after integrating over the internal directions associated to the torus,

\begin{equation}\label{actionfirstorder}
S=\frac{g_{s}^{2}}{16\pi G_{\rm N}^{(6)}}\int d^{6}x\sqrt{|g|}\, e^{-2 \phi}\left( R-4(\partial \phi)^2+\frac{1}{12}H^2+\frac{\alpha'}{8}R_{(-)}{}_{\mu\nu\rho\sigma}R_{(-)}{}^{\mu\nu\rho\sigma}\right)\, ,
\end{equation}

\noindent
where $G_{\rm N}^{(6)}=G_{\rm N}^{(10)}\left(2\pi \ell_{s}\right)^{-4}$. It is well-known that the Chern-Simons term in the local definition of $H$, eq.~\eqref{def:kalb-ramond}, hampers the application of the entropy function formalism, as this field depends on the curvature. Fortunately, at least in the cases of interest to us, it is possible to deal with this problem, see for instance \cite{Sahoo:2006pm, Sen:2007qy, Prester:2008iu, Faedo:2019xii}. Let us note, nevertheless, that the application of Wald's formula to theories that contain fields that do not transform as tensors is not justified in terms of the first law of thermodynamics, and it would be interesting to develop a more rigorous treatment of the entropy function formalism in the light of Refs.~\cite{Elgood:2020nls, Elgood:2020mdx}.
The first step is to rewrite the action in terms of the dual $2$-form ${\tilde B}$ defined in \eqref{deftildeH}. To achieve this purpose, we add the following total derivative to the action \eqref{actionfirstorder}

\begin{equation}
\tilde S=S-\frac{g^2_{s}}{16\pi G_{\rm N}^{(6)}}\int \left(H-\frac{\alpha'}{4}\Omega^{\rm L}_{(-)}\right)\wedge \tilde H\, .
\end{equation}

\noindent
The variation of the action with respect to $\tilde B$ gives the Bianchi identity of $H$, and the variation with respect to $H$ gives \eqref{deftildeH}, which can be used to eliminate $H$ in terms of $\tilde H$ everywhere. As a result, the dependence on the Riemann tensor has been made explicit, although now we have to deal with the non-covariant form of the Lagrangian. The resulting action can be split in three contributions: 
\begin{equation}
\tilde S=\int d^{6}x \sqrt{| g|}\,({\tilde {\cal L}}_{1}+{\tilde {\cal L}}_{2}+{\tilde {\cal L}}_{3})\, ,
\end{equation}

\noindent
where
\begin{eqnarray} \nonumber
{\tilde {\cal L}}_{1}&=&\frac{g_{s}^{2}}{16\pi G_{\rm N}^{(6)}}\, \left[ e^{-2 \phi}\left(R-4(\partial \phi)^2\right)+\frac{e^{2\phi}}{12}{\tilde H}^2\right]\, ,\\
\nonumber\\ \nonumber
{\tilde {\cal L}}_{2}&=&\frac{g_{s}^{2}}{16\pi G_{\rm N}^{(6)}}e^{-2\phi}\frac{\alpha'}{8}R_{(-)}{}_{\mu\nu\rho\sigma}R_{(-)}{}^{\mu\nu\rho\sigma}\, , \\
\nonumber\\
{\tilde {\cal L}}_{3}&=&\frac{g_{s}^{2}}{16\pi G_{\rm N}^{(6)}}\frac{\alpha'}{4}\frac{\epsilon^{\mu_{1}\mu_{2}\mu_{3} \nu_{1}\nu_{2} \nu_{3}}}{\left(3!\right)^2 \sqrt{|g|}} \Omega^{\rm L}_{(-)}{}_{\mu_{1}\mu_{2}\mu_{3}} {\tilde H}_{\nu_{1}\nu_{2}\nu_{3}}\, .
\end{eqnarray}

\noindent
The only contribution that it is not manifestly covariant is the last one, ${\tilde {\mathcal L}}_{3}$, but under some assumptions a total derivative can be added to recast it in a manifestly covariant form.  We denote the resulting Lagrangian as $\sqrt{|g|}{\breve{\mathcal L}}_{3}=\sqrt{|g|}{\tilde{\mathcal L}}_{3}+$ total derivative. It will be the sum of two contributions  ${\breve{\mathcal L}}_{3}={\breve{\mathcal L}}'_{3}+{\breve{\mathcal L}}''_{3}$, corresponding to the following split of the Chern-Simons 3-form

\begin{equation} \label{eq:splitCS}
\Omega^{\rm L}_{(-)}={\mathscr A}+\Omega^{\rm L}\, ,
\end{equation}

\noindent
where 

\begin{equation}\label{expressionA}
{\mathscr A}=\frac{1}{2}d\left(\omega^{a}{}_{b}\wedge H^{b}{}_a\right)+\frac{1}{4}H^{a}{}_{b}\wedge D H^{b}{}_{a}-R^{a}{}_{b}\wedge H^{b}{}_{a}+\frac{1}{2}H^{a}{}_{b}\wedge H^{b}{}_c\wedge H^{c}{}_{a}\, .
\end{equation}

\noindent
and $\Omega^{\rm L}$ is the Chern-Simons 3-form associated the Levi-Civita spin connection $\omega^{a}{}_{b}$. The first contribution ${\breve{\mathcal L}}'_{3}$ is obtained from the first term in \eqref{eq:splitCS}, after adding a total derivative that cancells the one in \eqref{expressionA}, namely

\begin{equation}\label{eq:thirdcontribution}
\sqrt{|g|}{\breve {\cal L}}'_{3}=\frac{g_{s}^{2}}{16\pi G_{\rm N}^{(6)}}\frac{\alpha'}{4}\frac{\epsilon^{\mu_{1}\mu_{2}\mu_{3} \nu_{1}\nu_{2} \nu_{3}}}{(3!)^2 } {\tilde{\mathscr A}}{}_{\mu_{1}\mu_{2}\mu_{3}} {\tilde H}_{\nu_{1}\nu_{2} \nu_{3}}\, ,
\end{equation}

\noindent
with $\tilde{\mathscr A}={\mathscr A}-\frac{1}{2}d\left(\omega^{a}{}_{b}\wedge H^{b}{}_a\right)$.

 We are left with the second contribution due to $\Omega^{\rm L}$. For this we can use that, in the family of solutions considered, the six-dimensional metric \eqref{ansatz6d} is the sum of two three-dimensional metrics ---parametrized by the coordinates $\{t, r, z\}$ and $\{\theta, \phi, \eta\}$ respectively--- and that also ${\tilde H}$ is the sum of two contributions according to this splitting of the metric. Then, we have

\begin{equation}\label{eq:fourthcontribution}
\sqrt{|g|}{\tilde {\cal L}}''_{3}=\frac{g_{s}^{2}}{16\pi G_{\rm N}^{(6)}}\frac{\alpha'}{4}\frac{\epsilon^{\mu_{1}\mu_{2}\mu_{3} \nu_{1}\nu_{2} \nu_{3}}}{\left(3!\right)^2 } \Omega^{\rm L}{}_{\mu_{1}\mu_{2}\mu_{3}} {\tilde H}_{\nu_{1}\nu_{2}\nu_{3}}=\frac{g_{s}^{2}}{16\pi G_{\rm N}^{(6)}}\frac{\alpha'}{4}\left(\Omega^{\rm L}_{\theta \phi \eta}{\tilde H}_{trz}-\Omega^{\rm L}_{trz}{\tilde H}_{\theta\phi z}\right)\, ,
\end{equation}

\noindent
where we have chosen the orientation $\epsilon^{tr\theta\phi\eta z}=+1$. The last information we need in order to write this in a manifestly covariant form is that for three-dimensional metrics admitting a spacelike isometry,
\begin{equation}
ds^{2}_{(3)}=\lambda^{2}[{\mathfrak h}_{\alpha \beta} dx^{\alpha} dx^{\beta}-(dx^{\sharp}+V_{\alpha} dx^{\alpha})^2]\, , \hspace{1cm} \alpha, \beta=1,2.
\end{equation}

\noindent
the Chern-Simons 3-form $\Omega^{\rm L}$ is given by \cite{Guralnik:2003we}

\begin{equation}\label{eq:omegaL}
\Omega^{\rm L}_{12\sharp}= {\mathfrak R} (dV)_{12}-(dV)_{12}(dV)^{21}(dV)_{12}+ \partial_{[1}{\mathscr V}_{2]}\, ,
\end{equation}

\noindent
where $\mathfrak R$ is the Ricci scalar of the two-dimensional metric ${\mathfrak h}_{\alpha \beta}$ and ${\mathscr V}$ is a certain 1-form which involves the spin-connection associated to ${\mathfrak h}_{\alpha\beta}$.  Again, $\sqrt{|g|}{\breve{\cal L}}''_{3}$ is obtained by adding a total derivative to the action that cancels the last term in \eqref{eq:omegaL}.


\emph{\underline{Corrections to the entropy function}}. The most convenient way to find the corrections to the entropy function is to evaluate the six-dimensional Lagrangian on the ansatz \eqref{ansatz6d},

\begin{equation}
f(v_1,v_2,\vec u, e_i, p_i)= \int dz d\eta d\theta d\phi \left[ \sqrt{|g|}\left({\tilde {\cal L}}_1+{\tilde {\cal L}}_2+{\breve {\cal L}}'_3+{\breve {\cal L}}''_3\right)\right]_{\eqref{ansatz6d}}\, .
\end{equation}
 Therefore, the function $f$ will be now the sum of four contributions, $f=f_1+f_2+f'_{3}+f''_{3}$:

\begin{enumerate}
\item The first contribution is exactly the same as the one we obtained in \eqref{eqfleadingorder}.

\item The second contribution will not be displayed since we do not need it to compute the first-order corrections. This is due to the fact that the curvature tensor $R_{(-)}{}_{\mu\nu\rho\sigma}$ vanishes when evaluated at the extremum \eqref{extremum}. Then, this contribution must be at least of second order in $\alpha'$, so we can simply ignore it.

\item The third contribution \eqref{eq:thirdcontribution} is

\begin{equation}
f'_{3}=\frac{g_{s}^2\alpha'}{2G_{\rm N}^{(4)}}\left[\frac{u_t^2p_4^2}{u_s^3v_1^2} \left(u_t^2p_4^2+u_s^2\left(u_t^2e_1^2-v_1\right)\right)+\frac{u_u^2e_3^2}{u_s^3v_2^2}\left(u_u^2e_3^2+u_s^2\left(u_u^2p_2^2-v_2\right)\right)\right]\, .
\end{equation}

\item Finally, the last contribution \eqref{eq:fourthcontribution} is 

\begin{equation}
f''_{3}=\frac{g_{s}^2\alpha'}{2G_{\rm N}^{(4)}}   \left[\frac{e_3 p_2 u_{u}^2 \left(2 p_{2}^2 u_{u}^2-v_{2}\right)}{v_{2}^2}-\frac{e_{1} p_{4} u_{t}^2 \left(v_{1}-2 e_{1}^2 u_{t}^2\right)}{v_{1}^2}\right]\, .
\end{equation}
\end{enumerate}

It is straightforward to check that \eqref{ansatz6dext2} is also an extremum of the corrected entropy function, as expected. However, the relation between the electric charges carried by the lower-dimensional vector fields and the source parameters is no longer the one we found at zeroth order in $\alpha'$, eq.~\eqref{eq:sources}. Now, taking into account the corrections to $f$, we find 

\begin{equation}
\begin{aligned}
Q_{1}=&\frac{G_{\rm N}^{(4)}}{g^2_{s}}\frac{\partial f}{\partial e_{1}}\Bigg|_{\rm{ext}}=\frac{\alpha'^2}{4 R^2_{z}R_{\eta}}n\left(1+\frac{2}{NW}\right) \,, \\
Q_{3}=&\frac{G_{\rm N}^{(4)}}{g^2_{s}}\frac{\partial f}{\partial e_{3}}\Bigg|_{\rm{ext}}= \frac{\alpha'}{4 R_{\eta}}\left(N-\frac{2}{W}\right)\, ,
\end{aligned}
\end{equation}

\noindent
which agree with the value of the Maxwell charges obtained in the previous subsection. Finally, evaluating the corrected entropy function $\cal E$ at the extremum, we get Wald's entropy

\begin{equation}
S_{\rm W 4d}=2\pi\sqrt{nwNW} \left(1+\frac{2}{NW}\right)=2\pi \sqrt{{\cal Q}_{+}{\cal Q}_{-}\left({\cal Q}_{0}{\cal Q}_{H}+4\right)}\, ,
\end{equation}

\noindent
as previously reported in \cite{Sahoo:2006pm, Prester:2008iu, Faedo:2019xii}.

\emph{\underline{Five-dimensional three-charge black holes}}. The above steps can be repeated to obtain this near-horizon solution after the following modifications are taken into account. In first place, since there are now less independent parameters, the appropriate ansatz is 

\begin{equation}\label{ansatz6d3c}
\begin{aligned}
d{\hat s}^2=\, &v_{1}\left(r^2dt^2-\frac{dr^2}{r^2}\right)-v_{2}\left(d\theta^2+\sin^2\theta d\phi^2\right)-u^{2}_{t}\left(dz-2e_{1}r dt\right)^2\\
&-v_2\left(d\psi+\cos\theta d\phi\right)^2\, ,\\
{\tilde H}=\,&2e_{3}\,dt\wedge dr\wedge dz+2p_{4}\,\sin\theta d\theta \wedge d\eta\wedge d\phi\, ,\\
e^{2\phi}=\, &\frac{u_{t}\sqrt{v_2}}{ u_{s}} \, .
\end{aligned}
\end{equation}

\noindent
Additionally, the first term in the right-hand-side of \eqref{eq:fourthcontribution} needs to be set to zero. The reason is the following. First, we notice that this term only depends on spatial components of the Riemann tensor, hence it cannot play a role in the computation of the entropy from Wald's formula. Second, the Chern-Simons 3-form of a 3-sphere is zero when evaluated using the Christoffel symbols, while reduces to a total derivative when evaluated using the spin connection. Hence, the inclusion of this term depends on the boundary conditions of the configuration. This leads us to the third and last consideration; from the structure of \eqref{eq:fourthcontribution}, it is clear that this term has the interpretation of magnetic source of the Kalb-Ramond field (or electric source of $\tilde{H}$) produced by the geometry of the Gibbons-Hawking space. In the four-dimensional solution, this term is responsible for a factor of $-1/W$ in the screening of the S5-brane charge (the other $-1/W$ comes from \eqref{eq:thirdcontribution}) produced by the KK gravitational instanton ---more details about this are given in the following section. Since in the five-dimensional solution the KK instanton number is zero, there can be no contribution from this term. 

Once these observations are considered, it is straightforward to check that \eqref{ansatz6dext2} with $W=1$ (which in this case does not have the physical interpretation of a charge, just like it does not indicate the presence of a KK monopole) gives again an extremum of the entropy function. Its evaluation gives

\begin{equation}
S_{\rm W 5d}=2\pi\sqrt{nwN} \left(1+\frac{2}{N}\right)=2\pi \sqrt{{\cal Q}_{+}{\cal Q}_{-}\left({\cal Q}_{0}+3\right)}\, .
\end{equation}

\noindent
For the charges, one gets

\begin{equation}
Q_{1}=\frac{\alpha'^2}{4 R^2_{z}}n\left(1+\frac{2}{N}\right) \,, \qquad
Q_{3}= \frac{\alpha'}{4 }\left(N-1\right)\, , \qquad
P_4= \frac{\alpha' w}{4} \, .
\end{equation}

\section{Kaluza-Klein monopoles and solitonic 5-branes}\label{sec:KKS5}
\subsection{General KK monopoles}
\label{sec:KK}

In the previous section we have described two families of regular black-hole solutions, with four and five non-compact dimensions respectively. Before describing the singular case of small black holes made by strings and momentum, it is convenient to study first the system formed by Kaluza-Klein monopoles and solitonic 5-branes.

At zeroth order, the unit charge Kaluza-Klein monopole is a well-known solution of string theory in which all fields are trivial except for the metric, that reads

\begin{eqnarray} \nonumber
ds^2 &=& dt^2-dz^\alpha dz^\alpha - {\cal H}^{-1}\left(d\eta+\chi\right)^2-{\cal H}\,\left(dr^2+r^2d\Omega^2_{(2)} \right) \, , \\ \label{eq:KKzero}
{\cal H}&=&1+\frac{R_\eta}{2r} \, , \qquad \chi=\frac{R_\eta}{2} \cos\theta d\varphi \, , \qquad
\eta \sim \eta+2\pi R_\eta \, .
\end{eqnarray}

\noindent
It is straightforward to check that this solution can be obtained from the family considered in section \ref{sec:zeroBH} setting $n=w=N=0$, $W=1$. On the other hand, at first sight it might not be obvious that the $\alpha'$-corrected Kaluza-Klein monopole is not automatically obtained performing the same operation on the solution described in section \ref{sec:firstBH}. There are several reasons why such procedure fails. In first place, it is unclear how to treat the $n/N \rightarrow 0/0$ indeterminacy that appears in the $\alpha'$-correction to $ \mathcal{Z}_{+} $, see \eqref{eq:exp_Z+2}. More importantly, the term $F(r;q_0)$ in $ \mathcal{Z}_{0} $ collapses into a harmonic pole that causes, among other effects, a divergence in the dilaton. On the other hand, the direct computation of the corrections to the original background \eqref{eq:KKzero} gives a regular configuration

\begin{eqnarray} \nonumber
ds^2 &=& dt^2-dz^\alpha dz^\alpha - \mathcal{Z}_0 \left[ {\cal H}^{-1}\left(d\eta+\chi \right)^2+{\cal H}\,\left(dr^2+r^2d\Omega^2_{(2)} \right) \right]\, , \\ \label{eq:KKfirst}
e^{2\phi}&=&e^{2\phi_\infty} \mathcal{Z}_0 \, , \qquad 
H=\star_\sigma d\mathcal{Z}_0 \, , \qquad \text{with} \, \, \, \, \,
\mathcal{Z}_0=1-\alpha' F(r;\tfrac{R_\eta}{2}) \, ,
\end{eqnarray}

\noindent
with $\mathcal{H}$ and $\chi$ unchanged. Recall that the function $F(r; q)$ is given by (in the case we are now considering, $W=1$)

\begin{equation}
F(r;q)=\frac{\left(r+q_{H}\right)\left(r+2q\right)+q^{2}}{4q_{H}\left(r+q_{H}\right)\left(r+q\right)^{2}} \, , \qquad  \, \, \, \, \, q_H=\frac{R_\eta W}{2} \, .
\end{equation}

A relevant property of the $\alpha'$-corrected heterotic KK monopole is that it carries -1 units of solitonic 5-brane charge, as defined in \eqref{eq:MaxwellS5charge}. This observation dates back to \cite{Sen:1997zb}, that arrived to this conclusion without explicitly finding \eqref{eq:KKfirst}, but using the fact that the KK monopole is a gravitational instanton with unit instanton number. The argument goes as follows. The Kalb-Ramond Bianchi identity has the form

\begin{equation}
dH=\frac{\alpha'}{4} R_{(-)}\,^a\,_b \wedge R_{(-)}\,^b\,_a \, .
\end{equation}

\noindent
In absence of matter fields at zeroth order, the right hand side is proportional to the Pontryagin density. Hence, upon integrating this equation over a four-dimensional Riemannian space, we obtain that the magnetic charge carried by $H$ is proportional to the gravitational instanton number. Working out the details, the aforementioned factor of $-1$ is obtained. The value can be read in a fast inspection of the asymptotic behaviour of $\lim_{r\rightarrow \infty}\mathcal{Z}_0=1-\frac{\alpha'}{r}+\dots$.

As emphasized by Sen in \cite{Sen:1997zb}, the fact that the KK monopole carries this S5 charge is a necessary condition for the consistency of S-duality of heterotic string theory. Thus, in order to properly understand the corrections to this string theory system, the perturbative solution must be constructed fixing the sources at $r=0$, while the asymptotic charges are allowed to vary. Observe that it is the asymptotic Maxwell charge of the Kalb-Ramond field strength the one that contains the information about the microscopic S5 charge (the S5 brane source charge vanishes for this configuration, $N=0$). Additionally, one notices that the truncation of sources directly in $\alpha'$-corrected solutions can produce a wrong answer; had we simply set $n=w=N=0$, $W=1$ directly in the general corrected solution of section \ref{sec:zeroBH}, we would not have obtained the appropriate value of S5 charge. 

The previous discussion extends straighforwardly to a multicenter configuration of KK monopoles. In \eqref{eq:KKzero} we can use a multicenter harmonic function, ${\cal H}=1+\sum_{i=1}^m \frac{R_\eta}{2r_i}$, where $r_i$ represents now the three-dimensional Euclidean distance measured from some point $\vec{x}_i$, interpreted as the location of a monopole. Likewise, $\chi$ must be appropriately modified\footnote{Its expression is not important for our discussion and can be readily found in the literature.}. It is well-known that the resulting space is a regular gravitational instanton, with instanton number given by the number of poles of the harmonic function, $m$. From the previous argument, one concludes that the multicenter configuration carries $-m$ units of S5 charge. The backreacted solution is still of the form of \eqref{eq:KKfirst}, with the already mentioned multicenter expressions for ${\cal H}$, $\chi$ and with  

\begin{equation}
{\cal Z}_0=1-\frac{\alpha'}{4} \left[\sum_{i=1}^{m} \frac{2}{R_\eta r_i}-\frac{({\vec\nabla} {\cal H})^2 }{{\cal H}^3}\right]=1-\frac{\alpha'}{4} \left[\sum_{i=1}^{m} \frac{2}{R_\eta r_i}-\frac{ R_{\eta}^2\sum_{i,j} \frac{(\vec {x}-{\vec {x}}_i)\cdot (\vec {x}-{\vec {x}}_j) }{r_i^3r_j^3}}{4\left(1+\sum_{i} \frac{R_{\eta}}{2r_i}\right)^3}\right]\,.
\end{equation}

The most general configuration is that of multicenter KK monopoles, each with generic charge. So far in this section, we have restricted to unit charge monopoles by setting the coefficient of all harmonic poles to $R_\eta/2$. Together with the fact that the coordinate $\eta$ has period $2\pi R_\eta$, this ensures that the metric is locally flat at the centers $\vec{x}_i$. On the contrary, a monopole with general charge is obtained if the coefficient is $W_i R_\eta/2$, with $W_i$ a positive integer. The resulting space presents conical singularities at the centers whenever the charge is larger than one, as described after \eqref{eq:KKcenter}. In what follows, we offer a detailed computation of the instanton number when these defects are present.
The gravitational instanton number is defined as 

\begin{equation}
{\mathfrak{n}} =-\frac{1}{16\pi^2} \int_{\mathcal{M}} R^a\,_b \wedge R^b\,_a \, ,
\end{equation}
where $\mathcal{M}$ denotes the four extended dimensions where the metric is non-trivial. In fact, since the solution is purely four-dimensional and the instanton number is independent of conformal rescalings of the metric, we can just evaluate the integral above in the metric 

\begin{equation}
ds_{\mathcal{M}}^2=\mathcal{H}^{-1}(d\eta+\chi )^2+\mathcal{H}\left(dr^2+r^2d\Omega_{(2)}^2\right)\, .
\end{equation}

\noindent
For simplicity, let us perform the computation in the case of one center, so that 
\begin{equation}
{\cal H}=1+\frac{W}{r}\, ,  \quad \chi=W\cos\theta d\varphi\, ,\quad W=1,2,\ldots,
\end{equation}

\noindent
where we are setting units such that $R_{\eta}=2$. Now,  the curvature tensor is self-dual, the instanton number can be expressed as\footnote{The self-duality of the Riemann tensor also implies that the vanishing of the Ricci tensor, $R_{\mu\nu}=0$.}

\begin{equation}\label{nKK1}
\mathfrak{n}=\frac{1}{32\pi^2}\int_{\mathcal{M}}d^4x\sqrt{g}\mathcal{X}_{4}\, ,
\end{equation}

\noindent
where $\mathcal{X}_{4}=R_{\mu\nu\rho\sigma}R^{\mu\nu\rho\sigma}-4R_{\mu\nu}R^{\mu\nu}+R^2$ is the Gauss-Bonnet density. If $\mathcal{M}$ were a manifold, this would be nothing but its Euler characteristic, but in our case one has to be careful with this interpretation due to the presence of a conic defect at $r=0$. 
In order to perform the computation we may first split $\mathcal{M}$ in two regions $r>r_0$ and $r<r_0$, that we may call $\mathcal{M}_{\infty}$ and $\mathcal{M}_{0}$, respectively. The instanton number is then the sum of ``Euler characteristics''

\begin{equation}\label{nKK2}
\mathfrak{n}=\mathcal{X}(\mathcal{M}_0)+\mathcal{X}(\mathcal{M}_{\infty})\, ,
\end{equation}

\noindent
where now, since each part $\mathcal{M}_p$ has a boundary, we have to take into account the boundary terms:

\begin{equation}\label{ECbdry}
\mathcal{X}(\mathcal{M}_p)=\frac{1}{32\pi^2}\int_{\mathcal{M}_{p}}d^4x\sqrt{g}\mathcal{X}_{4}+\frac{3}{16\pi^2}\int_{\partial\mathcal{M}_{p}}d^3x\sqrt{h}\left(K^{[i}_{[i}\mathcal{R}^{jk]}_{jk]}-\frac{2}{3}K^{[i}_{[i}K^{j}_{j}K^{k]}_{k]}\right)\, .
\end{equation}

\noindent
Here $h_{ij}$ is the induced metric on the boundary $r=r_0$, $\mathcal{R}$ is the intrinsic curvature and $K$ is the extrinsic curvature, defined as

\begin{equation}
K_{ij}=\frac{1}{2}\mathcal{L}_{n}h_{ij}\, ,
\end{equation}

\noindent
where ${\cal L}_{n}$ is the Lie derivative with respect to the normal vector $n$. Since the normal vectors to $\mathcal{M}_{0}$ and $\mathcal{M}_{\infty}$ are opposite, it is obvious that in \req{nKK2} the boundary terms cancel out and one is left with the integration in the whole volume, hence recovering \req{nKK1}. 
The evaluation of $\mathcal{X}(\mathcal{M}_{\infty})$ is straightforward and it yields

\begin{equation}
\mathcal{X}(\mathcal{M}_{\infty})=-\frac{W^2 (4 r+W)}{(r+W)^4}\Bigg|_{r_{0}}^{\infty}-\frac{W^2 (4 r_0+W)}{(r_0+W)^4}=0\, .
\end{equation}
This actually follows from the fact that $\mathcal{M}_{\infty}$ is topologically $\mathbb{S}^{1}\times \mathbb{S}^2\times [0,1)$ and from the factorization property of the Euler characteristic. 

\begin{figure}[t!]
\centering
\includegraphics[width=0.5\textwidth]{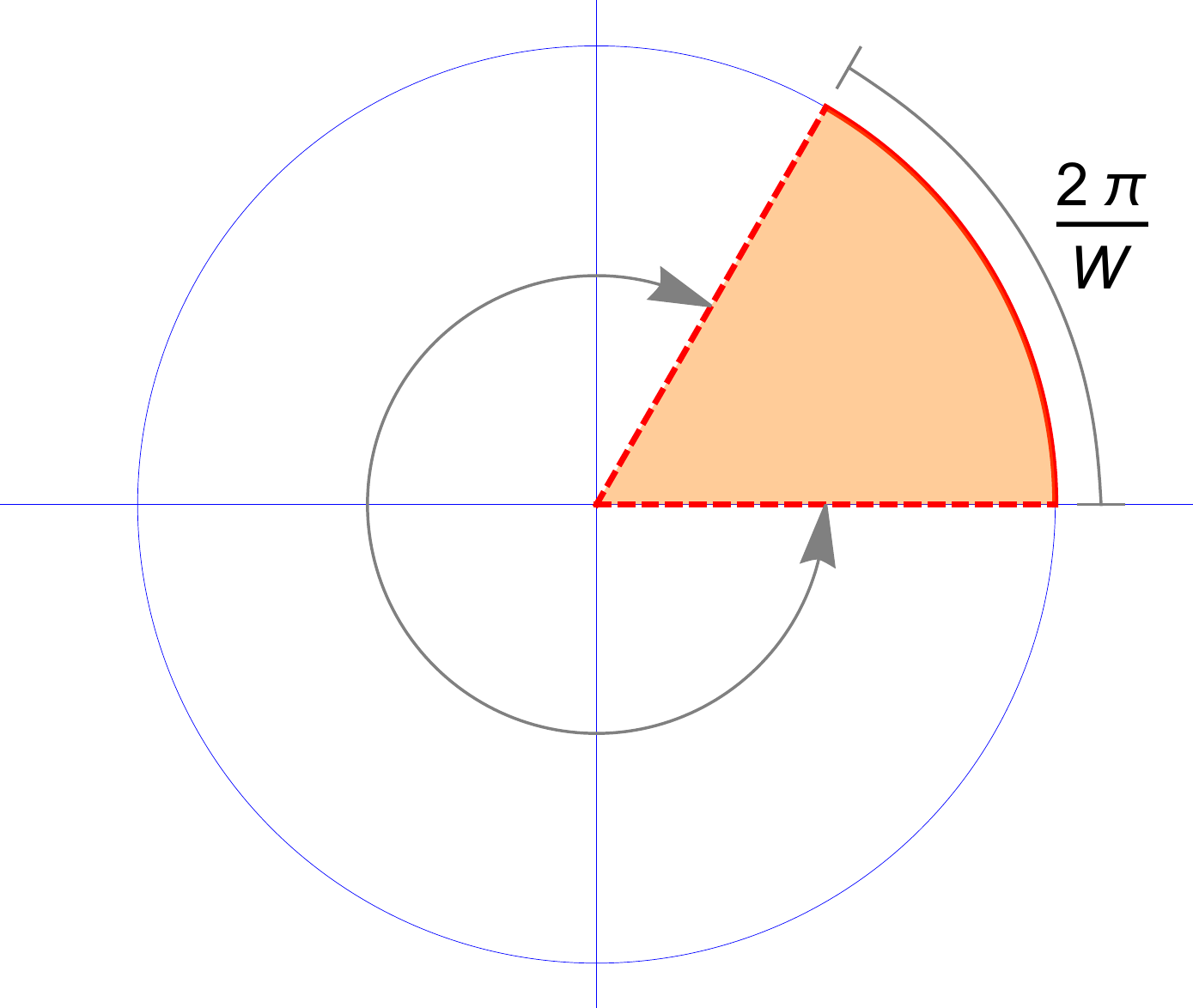}
\caption{Orbifold $B_4/\mathbb{Z}_{W}$. This is a slice of the unit ball in $\mathbb{E}^4$ where the sides (red dashed lines) are identified. The boundary is only composed of the arc of the circumference $r=1$ (red solid line).}
\label{fig:orbi}
\end{figure}

Let us now consider $\mathcal{M}_{0}$. We already mentioned that near $r=0$ the KK monopole becomes the orbifold $\mathbb{E}^4/\mathbb{Z}_{W}$. Therefore, $\mathcal{M}_{0}$ is is topologically $B_4/\mathbb{Z}_{W}$, this is, a $1/W$ slice of the unit ball in $\mathbb{E}^4$, with the sides identified as illustrated in Fig.~\ref{fig:orbi}.  
One can then apply \req{ECbdry} to this space in order to compute $\mathcal{X}(\mathcal{M}_0)$. Notice that, if the sides were not identified, one would need to take them into account in the boundary integral and there would be additional contributions coming from the vertices, so that the result would be $1$, \textit{i.e.}, the Euler characteristic of any simply-connected open set in $\mathbb{E}^4$. 
However, once they are identified they do not form part of the boundary and, for the same reason, there are no contributions from any of the vertices.  
Let us also stress that the curvature of this space is identically zero at all points, so that no bulk contribution can come from the cone at $r=0$ as well. 
Thus, the only contribution to \req{ECbdry} comes from the boundary at $r_0$. It is clear that adding up $W$ times that result one would get the corresponding value for the Euler characteristic of the disc, which is $1$. Therefore, we conclude that 
\begin{equation}
\mathfrak{n}=\mathcal{X}(\mathcal{M}_{0})=\frac{1}{W}.
\end{equation}
Note that this is, in fact, the \emph{orbifold} Euler characteristic of $\mathbb{E}^{4}/\mathbb{Z}_{W}$. Orbifold Euler numbers are naturally rational, and it has been known for long that the Gauss-Bonnet formula applied to orbifolds gives this result rather than the standard Euler characteristic \cite{satake1957}. Therefore, our fractional result for the instanton number of the higher-charge KK monopole should not come as a surprise. 

This result can be straightforwardly generalized to an arbitrary number of centers, in which case each center contributes as before and we get
\begin{equation} \label{eq:KKinstnum}
\mathfrak{n}=\sum_{i=1}^{m}\frac{1}{W_i}.
\end{equation}
Note that, once again, this is the orbifold Euler characteristic of the multicenter KK monopole.

While not a surprise, the fact that the result is a fractional number might feel uncomfortable when thinking about charge quantization. Additionally, there seems to exist a quite extended expectation\footnote{This is our personal perception of the issue, after having discussed about it with a respectable number of researchers. We do not know about any bibliographical support of this fact.} that the charges carried by $m$ unit charge KK monopoles should be the same than those carried by one monopole of charge $m$. While obviously the KK charges coincide, the former has $(-m)$ S5 charge while the latter has $(-1/m)$ if we use \eqref{eq:KKinstnum}. This could lead to the proposal that, in the presence of conical singularities, the S5 charge should not simply be the instanton number, but an additional contribution should be added. Such putative term should amount to $(W_i-1/W_i)$ for every defect, such that the S5 charge is always minus the KK charge. We do not know of any argument supporting the introduction of such term, and hence we will not do it here.

On the contrary, there are hints that the S5 charge might not necessarily be given in all cases by $-m$. On one side, let us note that the fractional KK contribution to the S5 charge is crucial in order to obtain the correct value for the black hole entropy in \eqref{eq:waldent5}, and in that case there is no conical singularity whatsoever. Besides this, we can find another possible argument in the study of the moduli space of the effective worldvolume theory of heterotic Kaluza-Klein monopoles, which was argued in \cite{Sen:1997js} to be that of BPS monopoles in a $SU(2)$ gauge theory\footnote{As described in \cite{Sen:1997js}, in M and type II theories the field content determines the effective theory of the KK worldvolume theories, that correspond respectively to $\mathcal{N}=1$ $(U(1)^m)$ gauge theories (M and type IIA) and $\mathcal{N}=(2,0)$ tensor multiplet (type IIB).}. The claim strongly relies on the fact that, after taking into account higher-curvature corrections, a collection of $m$ separated, unit charge monopoles has $(m,-m)$ KK and S5 charges, respectively. This is a unique feature of the heterotic theory. The conical singularity that appears in \eqref{eq:KKfirst} when several KK monopoles coincide is well understood in M and type IIA theories, where it produces an enhancement of the gauge symmetry group of the worldvolume theory, as well as in type IIB, where tensionless strings appear \cite{Sen:1997js}. On the other hand, a relevant property of the moduli space of BPS monopoles is that it has no singularities \cite{10.2307/j.ctt7zv206}. Hence, there would seem to be a contradiction between this fact and the possibility of having higher-charge heterotic KK monopoles with conical singularities. A way out of this problem would be that the S5 charge they carry is not the same as when they are separated, so that higher-charge KK monopoles have different quantum numbers than BPS monopoles.

\subsection{Adding S5 branes} \label{sec:KK+S5}

In previous section we have shown that higher-charge KK monopoles, if alone, have a discrete charge spectrum which does not obey standard quantization rules. A plausible interpretation could be that the corresponding solution to the equations of motion does not correspond to any actual state of string theory. Just like in classical electromagnetism, there are solutions to the field equations which are discarded once Dirac quantization is imposed, i.e. we shall only consider solutions in which the charge is an integer. In the case at hands, this implies that we need to add S5 branes. Remarkably, this addition also resolves the orbifold singularity and produces a regular, geodesically complete manifold. The solution has still the form given in \eqref{eq:KKfirst}, now with

\begin{equation}
\mathcal{Z}_0= 1+\frac{q_{0}}{r}- \alpha'\left[F(r;q_{0})+F(r;q_{H})\right]+\mathcal{O}(\alpha'{}^{2}) \, ,
\end{equation}

\noindent
where, we recall, $q_0=\tfrac{\alpha' N}{2R_\eta}$. Contrary to the situation in the previous section, the right solution is obtained if we truncate $n=w=0$ in the $\alpha'$-corrected black hole of section \ref{sec:firstBH}. Now, when we approach the $r \rightarrow 0$ region, we find that the orbifold singularity has been replaced by a semi-infinite cylinder smoothly glued to the Gibbons-Hawking space. Indeed, in this near-brane limit one finds

\begin{equation}
ds^2=dt^2-dz^\alpha dz^\alpha -\left( d\beta^2 + \alpha' NW d \Omega^2_{(3)/\mathbb{Z}_W} \right) \, , \qquad \phi=-\frac{\beta}{\sqrt{\alpha' NW}} \,. 
\end{equation}

\noindent
with $d \Omega^2_{(3)/\mathbb{Z}_2} $ the metric on the Lens space $S^3/\mathbb{Z}_W$, which has the form of the metric on the 3-sphere but its volume is only a $1/W$ fraction of it. The radial coordinate has been redefined as $\beta=\sqrt{\alpha' NW} \log r$, such that $r \rightarrow 0$ corresponds to an infinitely extended cylinder along $\beta \rightarrow -\infty$. For fixed values of $\beta$, $t$, $z^\alpha$, the geometry is that of a 3-sphere with points identified under the action of a discrete $\mathbb{Z}_W$ group that has no fixed points. 

The computation of the S5 charge, defined in \eqref{eq:MaxwellS5charge}, yields

\begin{equation}
\mathcal{Q}_0=N-\frac{2}{W} \, ,
\end{equation}

\noindent
which, in light of the discussion below eq.~\eqref{eq:cardyent}, we assume to be an integer. In particular, this means that the localized brane source charge $N$ can be fractional. This somewhat unusual value is a consequence of the $\mathbb{Z}_W$ quotient performed at the sphere that surrounds the brane. The topology of the space coincides with that of the previous subsection, where the conical singularity has been mapped to the asymptotic near-brane region $r\rightarrow 0$, and hence the KK gravitational instanton screens the S5 charge with a factor of $-1/W$. An additional factor of $-1/W$ comes from the introduction of the S5-brane localized sources. This latter effect, which has been mostly ignored in the literature, is a consequence of using the supersymmetric formulation of the heterotic theory given in \cite{Bergshoeff:1989de}, as described in \cite{Chimento:2018kop}. Recall that in this formulation the torsion component of the spin connection is $-\tfrac{1}{2} H_\mu \,^a \, _b dx^\mu$. Due to the presence of this term, a stack of S5 branes produces a new gravitational instanton that backreacts as a source in the $\alpha'$-corrected Bianchi identity. An elementary observation that, nevertheless, must be stressed is that, once the S5 branes are included, there is no uncertainty in the computation of the S5-brane charge, as the manifold has no singularity anymore. 

In summary, we have seen that a non-perturbative modification of the higher-charge KK monopole, that involves the introduction of S5 branes, allows to solve the problem of charge quantization and simultaneously resolves the conifold singularity. The most relevant effect of higher-curvature corrections is to modify the charges of the zeroth-order solution, which must be allowed to vary in the perturbative expansion. We have also seen that the truncation of charges in a general corrected solution may produce an incorrect result.

\section{Higher-curvature corrections to small black holes and rings}\label{sec:SBHs}

After having discussed those solutions made up of Kaluza-Klein monopoles and solitonic 5-branes, we now turn our attention into the study of small black holes and rings, consisting solely of  fundamental strings wrapping an compact direction (denoted by $z$) with momentum flowing along them.

Regarding small black holes, special attention has been paid to the four-dimensional ones. At leading order in the  $\alpha'$ expansion, they were shown to be solutions of the heterotic effective action  characterised by a singular horizon with vanishing area \cite{Sen:1994eb}. The inclusion of quadratic-curvature corrections was studied in detail in \cite{Cano:2018hut}, where it was found that they do not regularize the singular supergravity solution. A similar analysis was carried out for five-dimensional small black rings, obtaining analogous conclusions \cite{Ruiperez:2020qda}: small black rings in five dimensions are singular in the supergravity approximation and the $\alpha'$ corrections do not cure this behaviour. The aim of this section is to present a general treatment and extend these results to any number of dimensions. 

Let us begin with a discussion of the heterotic backgrounds constructed in the mid-1990s in \cite{Dabholkar:1995nc, Callan:1995hn} which describe heterotic strings carrying arbitrary right-moving momentum waves, generalizing those of \cite{Dabholkar:1990yf, Garfinkle:1992zj}. These solutions preserve half of the spacetime supersymmetries (see Appendix~\ref{app:SUSY} for further details) and their form is the following

\begin{equation}\label{eq:10dmetricans} 
\begin{aligned}
ds^2=\,&\frac{2}{\zm}du\left(dt+\omega-\frac{\zp}{2}du\right)-ds^2 (\mathbb{E}^{d-1})-d{\vec z}^{2}_{(9-d)} \ ,\\ 
B=\,&\zm^{-1} \, du\wedge \left(dt+\omega\right) \ ,\\
e^{-2\phi}=\,&e^{-2\phi_\infty}\zm\ ,
\end{aligned}
\end{equation}
\noindent
where $ds^2(\mathbb{E}^{d-1})$ represents the metric of ${\mathbb E}^{d-1}$ and 

\begin{equation}
\begin{aligned}
\zm=&1+\frac{q_{-}}{||{\vec x}-{\vec F}||^{d-3}}\, , \\
\zp=&1+\frac{q_{+}+q_{-}{\dot F}^m{\dot F}^m}{||{\vec x}-{\vec F}||^{d-3}}\, , \\
\omega_{m}=&\frac{q_{-}{\dot {F}^{m}}}{||{\vec x}-{\vec F}||^{d-3}}\, ,
\end{aligned}
\end{equation}

\noindent
where $\vec x\in {\mathbb E}^{d-1}$, $q_{\pm}$ are constants and $F^{m}=F^{m}(u)$ are arbitrary functions of $u=t-z$. Derivatives with respect to this coordinate are denoted with a dot. Finally, ${\vec z}_{(9-d)}$ represent the coordinates over which the solution has been smeared and parametrize a torus ${\mathbb T}^{9-d}$ without internal dynamics. The position of the string in the non-compact directions is determined parametrically by 

\begin{equation}
\vec x=\vec F(u)\, .
\end{equation}
\noindent
For this solution to represent a closed string, we must demand that $\vec F(u)$ is periodic. We denote the periodicity of this function by $\ell$, which does not necessarily coincide with the periodicity of the compact coordinate $z$. Instead, we allow the function $F$ to be multi-valued on ${\mathbb S}^{1}_{z}$. All we demand is that the string closes after a finite number of revolutions along $z$. Therefore, $\ell=2\pi R_{z} w$, where $w=1,2, \dots$ represents the winding number along $z$.

Following \cite{Lunin:2001fv}, we can further smear the solution over the compact coordinate $z$, which yields the following solution

\begin{equation}
\begin{aligned}
\zm=&1+\int_{0}^{\ell}\frac{q_{-}}{||{\vec x}-{\vec F}||^{d-3}}\, du\, , \\
\zp=&1+\int_{0}^{\ell}\frac{q_{+}+q_{-}{\dot {\vec F}}\cdot {\dot {\vec F}}}{||{\vec x}-{\vec F}||^{d-3}}\, du\, , \\
\omega_{m}=&\int_{0}^{\ell}\frac{q_{-}{\dot {F}^{m}}}{||{\vec x}-{\vec F}||^{d-3}}\,du\, ,
\label{eq:zetasmall}
\end{aligned}
\end{equation}

\noindent
which has no dependence on $u$ anymore. Hence, it can be dimensionally reduced (on ${\mathbb T}^{9-d}\times {\mathbb S}^1_z$) to $d$ dimensions through a standard Kaluza-Klein reduction. The lower-dimensional metric that one obtains, in the Einstein frame, is 

\begin{equation}
ds^{2}_{(d)}=\left({\cal Z}_{+}{\cal Z}_{-}\right)^{\frac{3-d}{d-2}}\left(dt+\omega\right)^2-\left({\cal Z}_{+}{\cal Z}_{-}\right)^{\frac{1}{d-2}}ds^2 (\mathbb{E}^{d-1})\, .
\label{eq:dmetric}
\end{equation}

As we will see next, the lower-dimensional solutions can describe small black holes and rings for particular choices of $\vec F(u)$. Before discussing these choices, we shall make use of the results of \cite{Ruiperez:2020qda}, where the first-order $\alpha'$ corrections to the above class of backgrounds were computed. The form of the corrected solution turns out to be the same, so no other field components are activated by the corrections. This is in fact a consequence of supersymmetry. As we show in Appendix~\ref{app:SUSY} (see also \cite{Papadopoulos_2009}), \eqref{eq:10dmetricans} is the most general field configuration that one can write down for the DH states. The curvature corrections only modify the form of the function $\zp$. Then, the corrected solution is  \eqref{eq:10dmetricans} with $\zm, \zp, \omega$ given in terms of the zeroth-order solution (which we now denote as $\{\zmz,\zpz, \omega\}$) by the following expressions.

\begin{eqnarray}
\label{eq:solzm} \nonumber
\zm&=&\zmz+ \mathcal{O}(\alpha'^2)\, , \\ \nonumber
\label{eq:solzp}
\zp&=&\zpz+\alpha' \frac{\Omega^{(0)}{}_{mn}\Omega^{(0)mn}-2 \partial_m \zpz \partial_m \zmz}{4\zmz}+ \mathcal{O}(\alpha'^2)\, ,\\
\label{eq:solom}
\omega&=&\omega^{(0)}+ \mathcal{O}(\alpha'^2)\,,
\end{eqnarray}

\noindent
where $\Omega=d\omega$.

We shall now examine the subsequent black hole and black ring solutions arising from consideration of two particular profile functions $\vec{F}$.

\subsection{Static fundamental strings}\label{sec:staticSBHs}

We start by considering a constant ${\vec F}$, which corresponds to a static fundamental string. Without loss of generality, we can always set ${\vec F}=0$ after an appropriate change of coordinates. Plugging this static ansatz into \eqref{eq:solzm}, we find that

\begin{eqnarray}
\label{eq:zpsmallbh} \nonumber
{\cal Z}_{+}&=&1+\frac{\tilde{q}_+}{\rho^{d-3}}-\frac{(3-d)^2\alpha'}{2}\frac{\tilde{q}_+ \tilde{q}_-}{\rho^{{d-1}}(\rho^{d-3}+\tilde{q}_-)}+{\cal O}\left(\alpha'^2\right)\, ,\\
\nonumber\\
\label{eq:zmsmallbh}
{\cal Z}_{-}&=&1+\frac{\tilde{q}_-}{\rho^{d-3}}+{\cal O}\left(\alpha'^2\right)\, ,
\end{eqnarray}

\noindent
where we have made the definitions $\tilde{q}_+=q_+ \ell$ , $\tilde{q}_-=q_- \ell$ and $\rho= || \vec{x}||$. The Killing vector $\partial_t$ is timelike for positive values of $\rho$, becoming null in the $\rho \rightarrow 0$ limit, where the $tt$ component of the metric \eqref{eq:dmetric} vanishes, thus signaling the presence of an event horizon at $\rho=0$ whose area is given by\footnote{Note  that in order to have a regular ($d$-dimensional) metric we must ensure that $\zp>0$ for $\rho \in {\mathbb R}^+$, which in turn requires $q_+ <-b(q_-;d)$, where $b(q_-;d)$ is a certain positive-definite function of $q_-$ and $d$ which was determined numerically in \cite{Cano:2018qev} for the particular case of $d=5$.}

\begin{equation}
A_{H}=\frac{(d-3)\,\pi^{\frac{d-1}{2}}}{\Gamma\left(\frac{d-1}{2}\right)}\sqrt{\displaystyle{-2\alpha' \tilde{q}_+ \tilde{q}_-}}\,, 
\end{equation}

\noindent
We see, as anticipated, that the area of the horizon vanishes if curvature corrections are ignored (setting $\alpha'\to0$), giving rise to a naked singularity.  Then, it may naively seem, if one just looks at the lower-dimensional metric, that the naked singularity is cloaked by a regular event horizon once the corrections are taken into account. However, this is not the complete story: it is not hard to see that the Kaluza-Klein scalar $k$ one gets upon compactification of the $z$-coordinate takes the form\footnote{Remember that the $(d+1)$-dimensional metric is expressed in the Einstein frame.}
\begin{equation}
k=k_{\infty}\frac{{\cal Z}^{1/2}_{+}}{{\cal Z}_{-}^{\frac{d-3}{2\left(d-1\right)}}}\, , \hspace{0.5cm}\Rightarrow \hspace{0.5cm} k(\rho \to0)\sim \rho^{-\frac{2(d-2)}{d-1}}\, .
\label{eq:kkbehaviour}
\end{equation}

\noindent
Hence, it diverges at the horizon. This tells us that we cannot trust the $d$-dimensional metric \eqref{eq:dmetric} as it has been obtained through a singular dimensional reduction. This singular behavior  can also be detected directly in ten dimensions, where  the divergence of the KK scalar is reflected in a divergence of the ten-dimensional Ricci scalar, whose explicit form is
\begin{equation}
R=-\frac{7(d-3)^2 (\tilde{q}_-)^2}{2 \rho^2(\tilde{q}_-+\rho^{d-3})^2}\,.
\end{equation}

\noindent

\subsection{Rotating fundamental strings}

\label{subsec:frs}

Now let us consider that the string has a non-trivial profile function $\vec{F}$. We shall restrict, as in the previous literature (see e.g. \cite{Dabholkar:1995nc,Lunin:2001fv, Dabholkar:2006za, Balasubramanian:2005qu}), to circular profiles of the form

\begin{equation}
F^1=R \cos \left ( 2 \pi \frac{{\mathscr W} u}{\ell} \right) \, , \quad F^2=R \sin \left (2 \pi \frac{{\mathscr W} u}{\ell} \right) \, , \quad F^3= \cdots =F^{d-1}=0\,.
\end{equation}

\noindent
Such a configuration corresponds to a string winding a 2-torus spanned by $z$  and the polar angle $\psi$ in the $x^1-x^2$ plane. More concretely, this yields a helix profile for the string, which swirls around the $z$-direction while turning round along the circle $(x^1)^2+(x^2)^2=R^2$. This radius $R$ can be related to the momentum carried by the string \cite{Lunin:2001fv} and $\mathscr W$ (not to be confused with the charge of the KK monopole, denoted by $W$ in the previous section) represents the number of times the string is wrapped along the $\psi$ direction. 

It was shown in \cite{Dabholkar:2006za} (see also \cite{Balasubramanian:2005qu}) that this configuration, when reduced to $d>4$ dimensions,  gives raise to small black rings with two monopole and one dipole charges which are singular at leading order in the $\alpha'$ expansion. We shall now investigate the effect of the first-order corrections on these solutions. To this aim, we first re-derive the explicit form of zeroth-order solution for the above circular profile which was presented in \cite{Dabholkar:2006za}. It is convenient  to  introduce the following set $\{\xi, \psi, \eta, \phi_1, \dots, \phi_ {d-4}\}$ of new coordinates

\begin{equation}
\begin{aligned}
x^1=&\xi \cos \psi\, , \quad x^2= \xi \sin \psi \, ,\\
x^3=& \eta \cos (\phi_1) \, ,\quad \dots \quad x^{d-1}=\eta \sin(\phi_1) \dots \sin(\phi_{d-4})\,.
\label{eq:changecoord}
\end{aligned}
\end{equation}

\noindent
After some routine computations, one finds 

\begin{eqnarray} \nonumber
\mathcal{Z}_{\pm}^{(0)}&=&1+ \frac{\tilde{q}_{\pm}}{(\xi^2+\eta^2+R^2)^{\frac{d-3}{2}}} \, \, {}_2F{}_1 \left ( \frac{d-3}{4}, \frac{d-1}{4}; 1; \frac{4 R^2 \xi^2}{(\xi^2+\eta^2+R^2)^2} \right )\,, \\ 
\omegaz &=&\frac{(d-3) \pi  \tilde{q}_-  {\mathscr W} R^2 \xi^2 }{\ell (\xi^2+\eta^2+R^2)^{\frac{d-1}{2}}}\, \, {}_2F{}_1 \left ( \frac{d-1}{4}, \frac{d+1}{4}; 2; \frac{4 R^2 \xi^2}{(\xi^2+\eta^2+R^2)^2} \right ) d \psi\,,
\end{eqnarray}

\noindent
where ${}_2 F{}_1 (a,b;c;z)$ denotes the hypergeometric function and where we have defined $\tilde{q}_-\equiv q_- \ell$ and $\tilde{q}_+\equiv q_+ \ell +4 \pi^2 {\mathscr W}^2 R^2 q_-/\ell$.  Note that these results are strictly equivalent to those presented at \cite{Dabholkar:2006za} after identifying their notation $\{f_f,f_p,A_m\}$ with our notation $\{\zmz,\zpz,-\omega_m\}$. Regarding future manipulations, it is convenient to rewrite it in terms of the so-called ring coordinates \cite{Emparan:2001wn}, which are defined as

\begin{equation}
\xi=\frac{\sqrt{y^2-1}}{x-y} R \, , \quad \eta= \frac{\sqrt{1-x^2}}{x-y} R\, ,
\label{eq:ringcoord}
\end{equation}

\noindent
and whose range is $-\infty \leq y \leq -1$ and $-1 \leq x \leq 1$.  The  metric \eqref{eq:dmetric} (of the zeroth-order solution) in this coordinates reads

\begin{equation}
\begin{aligned}
ds^2_{(d)}=(\zpz \zmz)^{\frac{3-d}{d-2}}(dt+\omegaz)^2- \frac{R^2 (\zpz \zmz)^{\frac{1}{d-2}}}{(x-y)^2} \bigg [&\frac{d y^2}{y^2-1}+(y^2-1) d \psi^2+\frac{d x^2}{1-x^2} \\
&+(1-x^2) d \Omega^2_{(d-4)} \bigg]\,,
\end{aligned}
\end{equation}

\noindent
where $d \Omega^2_{(d-4)}$ denotes the metric of ${\mathbb S}^{(d-4)}$ and 

\begin{eqnarray}
\label{eq:zpmzero} \nonumber
\mathcal{Z}_{\pm}^{(0)}&=&1+ \tilde{q}_{\pm} \left ( \frac{y-x}{2 R^2 y }  \right )^{\frac{d-3}{2}} \, \, {}_2F{}_1 \left ( \frac{d-3}{4}, \frac{d-1}{4}; 1; 1-\frac{1}{y^2} \right )\,, \\ 
\label{eq:omegazeroexp}
\omegaz &=&\frac{(d-3) \pi  \tilde{q}_- {\mathscr W} R^4  (y^2-1) }{\ell (x-y)^2} \left ( \frac{y-x}{2 R^2 y } \right )^{\frac{d-1}{2}}  {}_2F{}_1 \left ( \frac{d-1}{4}, \frac{d+1}{4}; 2; 1-\frac{1}{y^2} \right ) d \psi\,.
\end{eqnarray}

\noindent
It is not difficult to see that the norm of the Killing vector $\partial_{t}$ vanishes at $y\to-\infty$.  However, this does not correspond to a regular horizon since this null hypersurface has vanishing area and, furthermore, the curvature blows up there, exactly what one finds for the static small black holes discussed in the previous subsection.

Let us then take into account the corrections. Given this zeroth-order solution, it is straightforward to use \eqref{eq:solzp} to find the corrections to $\zp$. Since its explicit form involves long expressions which are not particularly illuminating, we relegate it to Appendix~\ref{app:corrections}, see eq.~\eqref{eq:zpcorrected}. It suffices to know that the near-horizon behavior of the function $\zp$ is modified by the $\alpha'$ corrections as

\begin{equation}
\zp \underset{y \rightarrow -\infty}{\sim}  \vert y \vert ^{d-4}+\alpha' \vert y \vert ^{d-2}\, ,
\end{equation}

\noindent
while 

\begin{equation}
\zm \underset{y \rightarrow -\infty}{\sim} \vert y \vert^{d-4}\, , \hspace{1cm}\omega \underset{y \rightarrow -\infty}{\sim} \vert y \vert^{d-4} d \psi\,.
\end{equation}

\noindent 
Then, we find that the area of the would-be horizon scales as
\begin{equation}
\label{eq:horarescaling}
A_{\mathrm{H}}  \sim\lim_{y \rightarrow -\infty} \left[\left ( \frac{(\zp \zm)^{\frac{1}{d-2}} R^2}{y^2} \right ) ^{\frac{d-2}{2}} y \right]\sim \sqrt{\alpha' \left (\ell^2 \tilde{q}_+-4 \pi^2 \tilde{q}_- R^2 \mathscr{W}^2\right )} \sim \sqrt{n w -J \mathscr{W}}\,,
\end{equation}
where $n$ represents the units of momentum carried by the fundamental string and $J$ its angular momentum. The last expression is obtained upon use of eqs. (A.4) and (A.13) of \cite{Dabholkar:2006za}, which relate the parameters $\tilde{q}_\pm$ and $J$ with $n, w$ and $\mathscr{W}$ as follows:
\begin{equation}
\frac{\tilde{q}_-}{16 \pi G_N^{(d)}}=\frac{\Gamma \left ( \frac{d-1}{2} \right ) }{2(d-3) \pi^{\frac{d-1}{2}}} \frac{R_z w}{\alpha'}\, , \quad \frac{\tilde{q}_+}{16 \pi G_N^{(d)}}=\frac{\Gamma \left ( \frac{d-1}{2} \right ) }{2(d-3) \pi^{\frac{d-1}{2}}} \frac{n}{R_z}\, , \quad J=\frac{R^2 \mathscr{W}}{\alpha'}\,.
\end{equation}
Note that the result \eqref{eq:horarescaling} is in agreement with the scaling argument of \cite{Sen:1996pb, Dabholkar:2006za}. 

In another vein, we check that \eqref{eq:horarescaling} vanishes at leading order (as we anticipated) while receiving a finite correction once the first-order $\alpha'$ corrections are included. However, one should be aware of the fact that this finiteness is only a mirage, and it actually comes from the combination of two divergences. In order to see this explicitly, let us first carry out the change of coordinates \cite{Emparan:2006mm}

\begin{equation}
r=-\frac{R}{y}\, , \quad x= \cos \theta\,,
\label{eq:newcoord}
\end{equation}

\noindent
which maps the $y\to -\infty$ hypersurface to the $r\to0^+$ hypersurface. Using this coordinates, our metric \eqref{eq:dmetric} reads

\begin{equation}
ds^2_{(d)}=(\zp \zm)^{\frac{3-d}{d-2}}(dt^2+\omega)^2- \frac{(\zp \zm)^{\frac{1}{d-2}}}{(1+\frac{r \cos \theta}{R})^2} \bigg [\left ( 1-\frac{r^2}{R^2} \right )R^2 d \psi^2+\frac{dr^2}{1-\frac{r^2}{R^2}}+r^2d \Omega^2_{(d-3)} \bigg]\,,
\end{equation}

\noindent
where $d\Omega^2_{(d-3)}=d\theta^2+\sin^2\theta d\Omega^2_{(d-4)}$ and where 
it is assumed that $\zp, \zm$ and $\omega$ are expressed in terms of the new coordinates \eqref{eq:newcoord}. Surfaces of constant $r$ have topology $\mathbb{S}^1 \times \mathbb{S}^{d-3}$, where the $\mathbb{S}^1$ is charted by the $\psi$-coordinate. By using the near-horizon behavior of $\zp, \zm$ and $\omega$, we find that the radii $R_\psi$ and $R_{d-3}$ associated to $\mathbb{S}^1$ and $\mathbb{S}^{d-3}$ scale near the horizon as

\begin{equation}
R_\psi \underset{r \rightarrow 0^+}{\sim} r^{\frac{3-d}{d-2}} \sim \vert y \vert^{\frac{d-3}{d-2}}\, , \quad R_{d-3} \underset{r \rightarrow 0^+}{\sim} r^{\frac{1}{d-2}} \sim \vert y \vert^{-\frac{1}{d-2}}\,.
\end{equation}

\noindent
Consequently, the radius $R_{d-3}$ vanishes in the horizon while $R_\psi$ diverges. However, $R_\psi (R_{d-3})^{d-3}$, which is proportional to the area, is indeed finite in this limit, what justifies why the expression for the area does not diverge. 

This unusual behaviour of $R_\psi $ and $R_{d-3}$ clearly indicates that the metric \eqref{eq:dmetric} is singular at $y\to-\infty$ ($r\to0^+$). We can additionally check the existence of such singularity by computing its Ricci scalar $R_{d}$. Indeed, at zeroth order the Ricci scalar already diverges as $ R_d \underset{y \rightarrow -\infty}{\sim} \vert y \vert^{\frac{4}{d-2}}$, and after including the first order-corrections such behaviour is not regularized, since we find that $R_d \underset{y \rightarrow -\infty}{\sim}  (\alpha')^{-\frac{1}{d-2}} \vert y \vert^{\frac{2}{d-2}}$. This signals the persistence of the singularity, as well as the breakdown of the perturbative expansion.

\section{Fake small black holes}\label{sec:fakesmall}

The conclusion that one extracts from the previous section is that small black holes are not regularized by higher-curvature corrections. On the other hand, previous results in the literature have described the existence of regular black holes with a reduced number of charges when the curvature corrections are included, while at zeroth order the solutions with these reduced number of charges are singular. The possible compatibility of these seemingly contradictory facts is studied below. 

\subsection{Fake small black holes in four and five dimensions}

Let us recall some of the results described in section \ref{sec:firstBH}. The Wald entropy of the four-dimensional black hole in terms of the asymptotic charges has the following expression,

\begin{equation}
S_{\rm W4d}=2\pi \sqrt{{\cal Q}_{+}{\cal Q}_{-}\left({\cal Q}_{0}{\cal Q}_{H}+4\right)}\, .
\end{equation}

\noindent
Looking at this formula only, one sees that if any of the solitonic 5-brane or the Kaluza-Klein charges is set to zero, we obtain

\begin{equation}\label{eq:entropyfake4}
S_{\rm W4d}|_{{\cal Q}_{0}{\cal Q}_{H}=0} \stackrel{?}{=} 4\pi \sqrt{{\cal Q}_{+}{\cal Q}_{-}}\, ,
\end{equation}

\noindent
whose expression coincides with the microscopic degeneracy of the DH system \eqref{eq:smallentropy}. If it were possible to truncate both of the two charges in a consistent manner and, particularly, such that these expressions hold, this could be interpreted as a resolution of the horizon of four-dimensional small black holes via higher-curvature terms. 

However, in previous sections of the paper, we have illustrated how the truncation (or addition) of sources (and, consequently, of charges) in a particular solution is a procedure that needs to be handled with care. Indeed, if we remove the KK monopole from the general four-dimensional solution described in section \ref{sec:BHreview}, the result will be a singular space. The corrections to the functions ${\cal Z}_{0}$ and ${\cal Z}_{+}$ ---see eqs.~\eqref{eq:exp_Z+2}--- diverge when the KK-monopole charge vanishes, which tells us that this limit must be taken before computing the $\alpha'$ corrections. Doing so, one finds that the functions that determine the solution are given by 

\begin{eqnarray} \nonumber
{\cal Z}_{+}&=&1+\frac{q_+}{r}-\frac{\alpha' q_+ q_-}{2r^2\left(r+q_0\right)\left(r+q_-\right)}+{\cal O}\left(\alpha'^2\right)\, , \\ \nonumber
{\cal Z}_{0}&=&1+\frac{q_0}{r}-\frac{\alpha' q_0^2 }{4r^2\left(r+q_0\right)^2}+{\cal O}\left(\alpha'^2\right)\, , \\ \nonumber
{\cal Z}_{-}&=&1+\frac{q_-}{r}+{\cal O}\left(\alpha'^2\right)\, , \\ \label{eq:noKK}
{\cal H}&=&1+{\cal O}\left(\alpha'^2\right)\, ,
\end{eqnarray}

\noindent
which yield singularities in the spacetime and matter fields. We notice that if one further truncates the S5-brane charge, which here is achieved by imposing $q_{0}=0$, one recovers the (singular) solution derived in section~\ref{sec:staticSBHs} for the particular case of $d=4$. Clearly, the formula \eqref{eq:entropyfake4} is not correct for the resulting configuration.

On the other hand, we recall that, as outlined in section~\ref{sec:firstBH}, the vanishing of the Maxwell S5-brane charge does not necessarily imply the absence of S5-branes when KK monopoles are present, as we can have contributions from the higher-curvature terms in the Bianchi identity. Concretely, for the four-dimensional system we have that ${\cal Q}_{0}=0$ if the following relation between the sources holds,

\begin{equation}
NW=2\, .
\end{equation}

\noindent
As a result, we get a black hole with a regular horizon and whose S5-brane charge is completely screened. When a black hole has this property, we call it a \emph{fake small black hole}. In this case, the expression for the entropy in \eqref{eq:entropyfake4} is correct, and its value coincides with that of the DH states, $4\pi \sqrt{{\cal Q}_{+}{\cal Q}_{-}}$. However, since the KK monopole charge is necessarily non-vanishing (otherwise, the functions would be given by \eqref{eq:noKK}), the solution cannot be interpreted as a small black hole with regular horizon. Indeed, it is an \emph{ordinary} black hole with four type of sources which is already regular at zeroth-order in $\alpha'$, with the special property that its S5 brane charge is screened by the higher-curvature corrections. Additionally, we point out that the presence of S5-branes, even if its charge is screened, influences the amount of supersymmetry preserved by these solutions, which is $1/4$ instead of the $1/2$ preserved by the DH states, see e.g.~\cite{Cano:2018qev, Cano:2018brq}. It is worth noticing that in this solution the Wald entropy satisfies $S_{\rm W 4d}=A/2G$, the same relation that was found for the solutions described in \cite{Dabholkar:2004dq, Sen:2004dp,  Hubeny:2004ji}.

A similar construction is also possible if there are five non-compact dimensions, in which case it is possible to have a regular horizon without KK monopole. If we set ${\cal Q}_{0}=0$ in the general solution of section \ref{sec:firstBH}, which according to \eqref{eq:S5charge5} implies $N=1$, we obtain a configuration with the same charges than the DH states. However, in this case the numerical factor of $4\pi$ in the DH entropy is not reproduced. Instead, upon use of \eqref{eq:waldent5}, one has

\begin{equation}
S_{\rm W5d}|_{{\cal Q}_{0}=0}=2\sqrt{3}\pi \sqrt{{\cal Q}_{+}{\cal Q}_{-}}\, .
\end{equation}

\noindent
Therefore, although there exists a regular five-dimensional fake small black hole ---that is, a black hole with only two asymptotic charges, ${\cal Q}_{+}$ and ${\cal Q}_{-}$---,  its entropy does not reproduce the degeneracy of the DH states. This mismatch is a natural consequence of the fact that the sources of this black hole include not only fundamental strings with momentum, but also one solitonic 5-brane. Once again, due to the presence of this brane, the amount of supersymmetry preserved by the five-dimensional fake small black hole differs from the DH states.


\subsection{Higher-dimensional small black holes and supersymmetry}

We have just seen that it is possible to have regular supersymmetric black holes with less than four (three) Maxwell charges in four (five) dimensions, provided these contain four (three) non-vanishing brane-source charges, which signals the presence of S5-branes and (in $4d$) KK-monopoles. 

The goal of this section is to show that it is not possible to have regular, supersymmetric near-horizon geometries in the heterotic theory compactified on ${\mathbb T}^{9-d}\times {\mathbb S}^1_{z}$ if $d\ge6$. When there are $d\ge6$ non-compact dimensions, the internal manifold is not large enough to allow for effective point-like sources in the non-compact directions coming from S5-branes and KK-monopoles wrapping the internal space. This would imply that only fundamental strings with momentum can act as effective point-like sources for these higher-dimensional solutions. Therefore, this result gives additional evidence that regularization of a singular horizon occurs through the (non-perturbative) introduction of sources, rather than higher-curvature corrections.



As in the rest of the manuscript, we assume that the torus ${\mathbb T}^{9-d}$ has trivial dynamics and that the near-horizon limit of the solutions enjoys a SO$(2,1)$ $\times$ SO$(d-1)$ symmetry. With these assumptions in mind, we proceed to write down the most general ansatz for the near-horizon geometry. For the sake of convenience, we use the $(d+1)$-dimensional fields,\footnote{i.e., we get rid of the torus ${\mathbb T}^{9-d}$, which does not play any rôle here.}

\begin{equation}\label{ansatzfakesbhs}
\begin{aligned}
d{s^2_{(d+1)}}=\,&v_{1}\left(r^2dt^2-\frac{dr^2}{r^2}\right)-v_{2}\,d\Omega^{2}_{(d-2)}-u^{2}_{k}\left(dz-2 e r dt\right)^2\, ,\\
e^{-2\phi}=&\frac{u_{k}}{u_{\phi}}\, , \hspace{1cm}\tilde H=p \,\omega_{{\mathbb S}^{d-2}}\, ,
\end{aligned}
\end{equation}

\noindent
where $\tilde H$ is the $(d-2)$-form field strength dual to the 3-form $H$, $\tilde H\equiv e^{-2\phi}\star_{(d+1)} H$, and $\omega_{{\mathbb S}^{d-2}}$ is the volume form of the round ${\mathbb S}^{d-2}$ sphere.\footnote{$\star_{(d+1)}$ denotes the $(d+1)$-dimensional Hodge star operator.}

By applying the entropy function formalism, one can check that there are no regular extrema of the entropy function at zeroth order in $\alpha'$. However, when higher-curvature corrections are taken into account, the system of algebraic equations that one has to solve becomes much harder to study. This is probably the reason why small black holes in $d\ge6$ dimensions have hardly been studied in the literature.

In spite of this, we can follow an alternative route to show that there are no regular, supersymmetric (small) black holes with the assumed SO$(2,1)$ $\times$ SO$(d-1)$ isometry in the near-horizon limit. The argument goes at follows. The dilatino Killing spinor equation is given by 

\begin{equation}
\left(\partial_{a}\phi \Gamma^{a}-\frac{1}{12}H_{abc}\Gamma^{abc}\right)\epsilon=0\, .
\end{equation}

\noindent
Since the dilaton is constant, the above equation reduces to $H_{abc}\Gamma^{abc}\epsilon=0$ and, for our ansatz, this equation can only be satisfied by a non-vanishing Killing spinor $\epsilon$ if $p=0$. Since $p$ is proportional to the winding charge of the fundamental string, this already proves that there are no regular, supersymmetric solutions describing the near-horizon of higher-dimensional small black holes. However, it is possible to go an step further and show that there are no regular supersymmetric solutions of this form at all. To do this, we can use the integrability condition of the gravitino Killing spinor equation, which reduces to

\begin{equation}
R_{abcd}\Gamma^{cd}\epsilon=0\, ,
\end{equation}

\noindent
since the 3-form field strength $H$ vanishes. Contracting this equation with $\Gamma^b$ and using eq.~(2.6) of \cite{Gran:2005wf}, we arrive to

\begin{equation}\label{eq:integrabilitycondreduced}
R_{ab}\Gamma^{b}\epsilon=0\, .
\end{equation}

\noindent
It is not difficult to see by explicitly computing the Ricci tensor of the metric  \eqref{ansatzfakesbhs} that the above integrability condition cannot be satisfied in our configuration for any choice of the parameters.

Let us explain why this argument only holds if $d\ge 6$.  Notice that if $d=5$, $\tilde H$ is a 3-form and therefore the ansatz \eqref{ansatzfakesbhs} is not the most general one, as one can write down an electric term for $\tilde H$ of the form

\begin{equation}
\tilde H=2\tilde e\, dt\wedge dr \wedge dz+ p\, \omega_{{\mathbb S}^{3}}\, ,
\end{equation}

\noindent
which precisely indicates the presence of S5-branes, as they are electric sources of the dual Kalb-Ramond field strength.  The argument works exactly in the same way for $d=4$ non-compact dimensions, see eq.~\eqref{ansatz6d}. At the extremum of the entropy function, $\tilde H$ turns out to be self-dual and the dilatino KSE is solved by a non-vanishing Killing spinor $\epsilon$ satisfying 

\begin{equation}
\left(1-\Gamma^{012345}\right)\epsilon=0\, .
\end{equation}

\section{Conclusions}

\label{sec:conc}

In this article, we have studied supergravity field configurations that can be interpreted as sourced by the presence of different combinations of fundamental heterotic strings (carrying winding and momentum), solitonic 5-branes and Kaluza-Klein monopoles at first order in the higher-curvature expansion and in several dimensions.
The most relevant effect produced by the higher-curvature corrections is to introduce non-linear couplings between fields, such that there are delocalized sources in some of the equations of motion. This produces a shift in the mass and in some of the charges of the solution, which can have a negative character. An interesting phenomenon is that, in few specific cases ($\mathcal{Q}_0=0$ in $d=5$ and $\mathcal{Q}_0=-1$ in $d=4$), the value of the charges does not uniquely determine the solution, and more information is needed for that purpose.

 Depending on which sources are present, the solutions describe a black hole, a soliton or a naked singularity. The inclusion of first order corrections in the higher-curvature expansion of the effective theory does not change the character of the solution, but the addition or truncation of sources may do it. This latter operation is intrinsically non-perturbative and modifies substantially the properties of the fields at zeroth-order in the expansion. We show that, for this reason, the computation of curvature corrections and the truncation of sources are two processes that do not always commute; starting from a zeroth-order solution, the same result is not necessarily obtained if the two operations are performed in different order. 

In addition, we have shown that one gets a consistent string theoretic interpretation of a perturbative solution when the sources are kept fixed in the higher-curvature expansion, allowing variations in the value of the charges. This plays a fundamental role in the study of Kaluza Klein monopoles (as first noticed more than 20 years ago in \cite{Sen:1997zb}) and small black holes. As a consequence of these observations, small (2-charge) black holes corresponding to the DH system remain singular when quadratic curvature corrections are included. 

On the other hand, we note that the corrections imply the existence of regular 3-charge black holes in four dimensions, and 2-charge ones in five, where the vanishing charge is that of the S5 brane. Since all 3- and 2-charge solutions are singular at zeroth order in $\alpha'$, it might seem that the corrections resolve the singularities. However, following our previous discussion, what really happens is that the system described by these solutions is already regular at zeroth-order, and the effect of the corrections is to screen the charge of the S5 brane, which has a localized source.  
In $d=4$, the corresponding 3-charge solution has the same entropy as the DH system, but neither the charges, the sources nor the supersymmetry are equal to those of a string carrying momentum, and hence we refer to it as a fake small black hole. The matching of the entropies is most likely a coincidence, since in the five-dimensional case the regular 2-charge solution does not reproduce the entropy of the DH system. In turn, there is another 2-charge solution with no localized sources of S5 branes which is singular  --- this is the one describing the DH system. Likewise, in higher dimensions all 2-charge solutions are singular due to the absence of S5 branes.




\section*{Acknowledgments}

We are thankful to Tomás Ortín for years of collaboration and guidance in this field. We thank Atish Dabholkar and Ashoke Sen for useful discussions. The work of PAC is supported by a postdoctoral fellowship from the Research Foundation - Flanders (FWO grant 12ZH121N). The work of \'AM is funded by the Spanish FPU Grant No. FPU17/04964. \'AM was further supported by the MCIU/AEI/FEDER UE grant PGC2018-095205-B-I00 and by the ``Centro de Excelencia Severo Ochoa'' Program grant SEV-2016-0597. The work of PFR is supported by the Alexander von Humboldt Stiftung. The work of AR is supported by the Department of Physics and Astronomy Galileo Galilei with funds of the project PRIN 2017 "Supersymmetry Breaking with Fields, Strings and Branes".

\appendix
\section{Supersymmetry analysis}\label{app:SUSY}

The main purpose of this appendix is to show that the ansatz used in Section \ref{sec:SBHs} to describe small black holes is the most general one with no dependence on the coordinate $u$ preserving half of the spacetime supersymmetries at first order in $\alpha'$. We shall make a wide use of the results of \cite{Fontanella:2019avn} but we also refer to \cite{Gran:2005wf, Gran:2007kh} where supersymmetric heterotic backgrounds have been studied and classified using different techniques as those employed in \cite{Fontanella:2019avn}. The results of this appendix are contained in the general classification of half-supersymmetric heterotic backgrounds of \cite{Papadopoulos_2009}. Here we offer a different re-derivation of some of the results contained in this reference.

\subsection{General form of supersymmetric configurations}
 
According to this reference, the metric of a supersymmetric configuration can always be written as
\begin{equation}
ds^2=2f\left(du+\beta\right)\left[dt+K \left(du+\beta\right)+\omega\right]-h_{\underline{mn}}\, dx^{m}dx^{n}\, ,
\end{equation}

\noindent
where $\omega=\omega_{\underline m}dx^{m}$ and $\beta=\beta_{\underline m} dx^m$ are 1-forms on the eight-dimensional space charted by the coordinates $x^{m}$ and $f$ and $K$ are functions defined on this manifold\footnote{In general, objects occurring in the metric may also depend on $u$. We assume no dependence on this coordinate in order to perform a standard KK reduction over this internal direction.}. It is convenient to introduce the following zehnbein basis 

\begin{equation}
e^{+}=f\left(du+\beta\right)\, , \hspace{0.5cm} e^{-}=dt+K \left(du+\beta\right)+\omega\, , \hspace{0.5cm} h_{\underline {mn}}\,dx^{m}dx^{n}=e^{m}e^{n}\, \delta_{mn}\, .
\end{equation}
The components of the spin connection $\omega^{ab}$ in this basis are\footnote{In our conventions, we have that $de^{a}=+\omega^{a}{}_{b}\wedge e^b$, with $a, b=+,-, m$.}
\begin{eqnarray}
\label{eq:spinconnection1}
\omega_{+-}&=&-\frac{1}{2}\partial_{m}\log f \, e^{m}\, ,\\
\omega_{+m}&=&-f^{-1}\partial_m K \, e^{+}-\frac{1}{2}\partial_{m}\log f \,e^-+\frac{1}{2}\left(K d\beta+d\omega\right)_{nm}\, e^n\, ,\\
\omega_{-m}&=&-\frac{1}{2}\partial_{m}\log f \,e^++\frac{f}{2}\left(d\beta\right)_{nm}\, e^n\, ,\\
\omega_{mn}&=&\frac{1}{2}\left(K d\beta+d\omega\right)_{mn} e^++\frac{f}{2}\left(d\beta\right)_{mn}\,e^{-}\,+\varpi_{pmn}\, e^{p} \,,
\end{eqnarray}
\noindent
where we have defined $\varpi_{mnp}$ to be the spin connection associated to $h_{\underline {mn}}$, which satisfies that $de^m=-\varpi_{mn}\wedge e^n$.

In order for a configuration to be supersymmetric, several conditions need to be accomplished. First, the torsionful spin connection $\Omega_{(+)}{}^a{}_{b}\equiv \omega^a{}_b+\frac{1}{2}H_{c}{}^a{}_b \, e^c$ must fulfil that\footnote{We note there is an error in eq.~(3.17) of \cite{Fontanella:2019avn} since $\Omega_{(+)}{}_{-mn}=H_{-mn}$, which is non-vanishing in general. Therefore, the only constraint on $\Omega_{(+)}{}_{-mn}$ comes from eq.~\eqref{selfdualityH_{-mn}}.} 
\begin{eqnarray}
\label{eq:Omega+1}
\Omega_{(+)}{}_{[ab]-}&=&0\, ,\\
\label{eq:Omega+2}
\Omega_{(+)}{}_{am-}&=&0\, ,\\
\label{eq:Omega+3}
\nabla_{(+)}{}_{a}\Omega_{mnpq}&=&0\, ,
\end{eqnarray}
\noindent
where $\Omega_{mnpq}$ is a 4-form which can be interpreted as a Spin(7) structure. As such, it possesses the following properties\footnote{We follow the convention of \cite{Fontanella:2019avn} and indices with same latin letter $m_{i}, n_{i}, \dots $ are totally antisymmetrized.}
  \begin{eqnarray}
   \label{eq:O1O1simple}
    \Omega_{m_{1}m_{2}m_{3}p}\Omega_{n_{1}n_{2}n_{3}p}
    & =&
      -9\Omega_{m_{1}m_{2}n_{1}n_{2}}\delta_{m_{3}n_{3}}
      +6\delta_{m_{1}m_{2}m_{3},\, n_{1}n_{2}n_{3}}\, ,
    \\
    \label{eq:O2O2noantisimplebis}
    \Omega_{m_{1}m_{2}p_{1}p_{2}} \Omega_{n_{1}n_{2}p_{1}p_{2}}
    & =&
      -4\Omega_{m_{1}m_{2}n_{1}n_{2}}
      +12\delta_{m_{1}m_{2},\, n_{1} n_{2}}\, ,
  \\
      \label{eq:O2O2antisimplebis}
    \Omega_{m_{1}n_{2}p_{1}p_{2}} \Omega_{n_{1}m_{2}p_{1}p_{2}}
   & =&
      +4\Omega_{m_{1}m_{2}n_{1}n_{2}}
      +6\delta_{m_{1}m_{2},\, n_{1} n_{2}}\, .
    \\
    \label{eq:O2O2bis}
    \Omega_{m_{1}m_{2}n_{1}n_{2}} \Omega_{m_{3}m_{4}n_{1}n_{2}}
    & =&
      -4 \Omega_{m_{1}\cdots m_{4}}\, ,
    \\
    \label{eq:O3O3}
    \Omega_{mp_{1}p_{2}p_{3}}\Omega_{np_{1}p_{2}p_{3}}
    & =&
    42\delta_{mn}\, ,
    \\
      \label{eq:O2bis}
    \Omega_{m_{1}\cdots m_{4}}  \Omega_{m_{1}\cdots m_{4}}
    & \equiv&
      \Omega^{2}
       = 14\cdot 4!\, .
\end{eqnarray}
Apart from eqs.~\eqref{eq:Omega+1}, \eqref{eq:Omega+2} and \eqref{eq:Omega+3}, there exists another set of conditions to be imposed. They constrain the components of $H_{abc}$ as follows:
\begin{eqnarray}
\label{selfdualityH_{-mn}}
H^{(-)}_{-mn}&=&0\, ,\\
\label{selfdualityH_{+mn}}
H^{(-)}_{+mn}&=&\frac{1}{48}\Omega_{m}{}^{s_1 s_2s_3}\nabla_{+}\Omega_{n s_1s_2s_3}\, , \\
\label{selfdualityH_{mnp}}
H^{(-)}_{mnp}&=&\frac{1}{7}\left(2\partial_{q}\phi-H_{+-q}\right)\Omega^{q}{}_{mnp}\, ,
\end{eqnarray}
\noindent
where we have made use of the projectors acting on 2-forms $\Theta_{mn}$ and 3-forms $\Psi_{mnp}$ defined in \cite{Fontanella:2019avn}: 
\begin{eqnarray}
\Theta_{mn}&=&\Theta^{(+)}_{mn}+\Theta^{(-)}_{mn}\, , \hspace{1cm} \Theta^{(\pm)}_{mn}=\Pi^{(\pm)}_{mnpq}\Theta_{pq}\, , \\
\Psi_{mnp}&=&
\Psi^{(+)}_{mnp}+
\Psi^{(-)}_{mnp}\, , \hspace{1cm} 
\Psi^{(\pm)}_{mnp}=\Pi^{(\pm)}_{mnpqrs}
\Psi_{qrs}\, , 
\end{eqnarray}
\noindent
where
\begin{eqnarray}
\Pi^{(+)}_{mnpq}&=&\frac{3}{4}\left(\delta_{mn, pq}+\frac{1}{6}\Omega_{mnpq}\right)\, , \\
\Pi^{(-)}_{mnpq}&=&\frac{1}{4}\left(\delta_{mn, pq}-\frac{1}{2}\Omega_{mnpq}\right)\, , \\
\Pi^{(+)}_{m_1m_2m_3n_1n_2n_3}&=&\frac{6}{7}\left(\delta_{m_1m_2m_3, n_1n_2n_3}+\frac{1}{4}\Omega_{m_1m_2n_1n_2}\delta_{m_3 n_3}\right)\, , \\
\Pi^{(-)}_{m_1m_2m_3n_1n_2n_3}&=&\frac{1}{7}\left(\delta_{m_1m_2m_3, n_1n_2n_3}-\frac{3}{2}\Omega_{m_1m_2n_1n_2}\delta_{m_3 n_3}\right)\, .
\end{eqnarray}
\noindent
Let us analyse all eqs. \eqref{eq:Omega+1}, \eqref{eq:Omega+2},  \eqref{eq:Omega+3},  \eqref{selfdualityH_{-mn}}, \eqref{selfdualityH_{+mn}} and \eqref{selfdualityH_{mnp}} carefully. First, we realize that conditions \eqref{eq:Omega+1} and \eqref{eq:Omega+2} tell us that all the components of $\Omega_{(+)}{}_{ab-}$ vanish\footnote{The component $\Omega_{(+)}{}_{++-}=\omega_{++-}$ is not fixed by these equations but it vanishes for the coordinates we have chosen, see \eqref{eq:spinconnection1}.}, implying that the components $H_{ab-}$ get fixed in terms of the objects that occur in the metric. We find 
\begin{equation}\label{eq:H_{abc}}
H_{m+-}=\partial_{m}\log f\, , \hspace{1cm} H_{mn-}=f \left(d\beta\right)_{mn}\, .
\end{equation}
Let us postpone the study of eq. \eqref{eq:Omega+3} for the moment. Regarding eq.~\eqref{selfdualityH_{-mn}}, we see it imposes that
\begin{equation}
\left(d\beta\right)^{(-)}_{mn}=0\, ,
\end{equation}
\noindent
so that the connection $\beta$ is that of an Abelian octonionic instanton \cite{Fubini:1985jm, Gunaydin:1995ku}. On the other hand, we check that eq.~\eqref{selfdualityH_{+mn}} can be rewritten by use of eqs.~\eqref{eq:O2O2noantisimplebis} and \eqref{eq:O3O3} as 
\begin{equation}
H^{(-)}_{+mn}=-2\, \omega^{(-)}_{+mn}=- (K d\beta+d\omega)^{(-)}_{mn}=-(d\omega)^{(-)}_{mn}\, .
\end{equation}
\noindent
Therefore $H_{+mn}$ can always be expressed as
\begin{equation}
H_{+mn}=H^{(+)}_{+mn}-(d\omega)^{(-)}_{mn}=-(d\omega)_{mn}+ K (d\beta)_{mn}+f^{-1}\xi_{mn}\, ,
\end{equation}
\noindent
for some two-form $\xi=\frac{1}{2}\xi_{mn}\, e^m \wedge e^n$ satisfying that $\xi^{(-)}_{mn}=0$. Consequently, the general form of $H$ for supersymmetric configurations with no dependence on $u$ is
\begin{equation}
\begin{aligned}
H=&d\log f\wedge e^+ \wedge e^-+fe^-\wedge d\beta+e^{+}\wedge\left(-d\omega+K d\beta+f^{-1}\xi\right)\\
&+\frac{1}{3!} H_{mnp}\, e^{m}\wedge e^{n}\wedge e^{p}\, ,
\end{aligned}
\end{equation}
\noindent
with $H_{mnp}$ satisfying \eqref{selfdualityH_{mnp}}, which can be rewritten by virtue of \eqref{eq:H_{abc}} as
\begin{equation}
H^{(-)}_{mnp}=\frac{1}{7}\partial_{q}\left(2\phi-\log f\right)\Omega^{q}{}_{mnp}\, .
\end{equation}
Now it is the moment to study condition \eqref{eq:Omega+3}. From the $a=\pm$ components of eq. \eqref{eq:Omega+3} and taking into account that $\Omega_{mnpq}$ is independent of $u$ and $t$, we find 

\begin{equation}
\Omega_{(+)}{}_{\pm [m| s}\Omega_{s |npq]}=0\, .
\end{equation}
\noindent
Contracting this equation with $\Omega^{rnpq}$, we arrive to the equivalent condition 
\begin{equation}
\Pi^{(-)}_{mnpq}\Omega_{(+)}{}_{\pm pq}=0\, ,
\end{equation}
\noindent 
which reduces to the self-duality conditions already derived for $d\beta$ and $\xi$. If instead we take $a=m$ at eq. \eqref{eq:Omega+3}, we deduce that $\Omega_{(+)mnp}$ has special holonomy $G \subseteq \mathrm{Spin}(7)$. This last condition can be expressed in a fairly compact way if one chooses a basis $\{e^{m}\}$ for which the components of $\Omega_{mnpq}$ are constant, which is known to always exist locally since $\Omega_{mnpq}$ defines a Spin(7) structure. In particular, in such a basis, we obtain the condition
\begin{equation}
\Pi^{(-)}_{mnrs}\Omega_{(+)}{}_{prs}=0\, .
\label{eq:Omega+puro+}
\end{equation}
\noindent


\subsection{Killing spinor equations}

Let us now study the Killing spinor equations (KSEs). It was proven in \cite{Fontanella:2019avn} that they are fulfilled by a constant spinor $\epsilon$ satisfying
\begin{eqnarray}
\label{firstconditionKS}
\Gamma^{+}\epsilon&=&0\, ,\\
\label{secondconditionKS}
\Pi^{(-)}\epsilon&=&0\, .
\end{eqnarray}
\noindent 
where 
\begin{equation}
\Pi^{(-)}=\frac{7}{8}\left(1-\frac{1}{336}\Omega_{m_{1}\dots m_{4}}\Gamma^{m_{1}\dots m_{4}}\right)\, .
\end{equation}
The first condition \eqref{firstconditionKS} annihilates half of the spacetime supersymmetries and, albeit in general half-supersymmetric configurations do not necessarily satisfy it, the class of half-supersymmetric solutions we are interested in (those describing superpositions of fundamental strings with momentum) does \cite{Papadopoulos_2009}. 
Hence, if we want to preserve exactly this amount of supersymmetry, we must find a way to avoid using \eqref{secondconditionKS} but still solving the dilatino and the gravitino KSEs. For that, we are going to impose extra conditions on the fields to ensure that such KSEs hold even if \eqref{secondconditionKS} does not. 

To this aim, we first concentrate on the dilatino KSE. It is convenient to use the rewriting of such KSE provided in eq. (3.47) of \cite{Fontanella:2019avn}:\footnote{We correct an innocent typo in eq. (3.47) of \cite{Fontanella:2019avn}.} 
\begin{equation}
\left[\frac{1}{2}\left(2\partial_{m}\phi-H_{+-m}\right)\Gamma^{m}-\frac{1}{12}\left(H_{mnp}\Gamma^{mnp}+3H_{-mn}\Gamma^{-}\Gamma^{mn} \right)\right]\Pi^{(-)}\epsilon=0\, ,
\end{equation}
\noindent
where we have already required \eqref{firstconditionKS}. Since we do not want to impose \eqref{secondconditionKS}, the term between brackets must vanish necessarily. Therefore,
\begin{equation}\label{conditionsdilatinoKSE}
\phi=\phi_{0}+\frac{1}{2}\log f\, , \hspace{1cm} H_{mnp}=0 \,, \hspace{1cm} \left(d\beta\right)_{mn}=0\, ,
\end{equation}
\noindent
where $\phi_{0}$ is an integration constant.

We now move to the gravitino KSE. For a constant spinor satisfying \eqref{firstconditionKS}, it is only necessary to check that 
\begin{equation}
\Omega_{(+)}{}_{amn}\Gamma^{mn}\epsilon=0\, .
\end{equation}
As we derived before at eq. \eqref{eq:Omega+puro+}, in a basis $\{e^{m}\}$ where the components of $\Omega_{m_{1}\dots m_{4}}$ are constant, we have that $\Omega^{(-)}_{(+)}{}_{amn}=0$. Therefore, we can use eq. (A.46a) of  \cite{Fontanella:2019avn} to show that 
\begin{equation}\label{gravitinoKSE}
\Omega_{(+)}{}_{amn}\Gamma^{mn}\epsilon=\Omega_{(+)}{}_{amn}\Gamma^{mn}\Pi^{(-)}\epsilon=0\, ,
\end{equation}
\noindent
which implies that either $\Omega_{(+)}{}_{amn}$ or $\Pi^{(-)}\epsilon\, $ must vanish. Since we do not want on any account to impose \eqref{secondconditionKS}, we require $\Omega_{(+)}{}_{amn}=0$ and this, in turn, demands
\begin{equation}
{\varpi}_{mnp}=0 \, , \hspace {1cm} \xi_{mn}=0\, ,
\end{equation}
\noindent
so that $e^m=dx^{m}$ and $h_{\underline{mn}}=\delta_{\underline{mn}}$. Consequently, the most general configuration preserving half of the spacetime supersymmetries with no dependence on $u$ is given by\footnote{The 1-form $\beta$ can always be removed via the coordinate transformation $u\rightarrow u-\chi$, where $d\chi=\beta$, since by \eqref{conditionsdilatinoKSE} $\beta$ is closed.}
\begin{eqnarray}
ds^2&=&2e^{2\left(\phi-\phi_{0}\right)}du\left(dt+K du+\omega\right)-dx^{m}dx^{m}\, , \\
H&=&2e^{2\left(\phi-\phi_{0}\right)}d\phi\wedge du \wedge (dt+\omega)-e^{2(\phi-\phi_0)} du\wedge d\omega\ .
\end{eqnarray}
\noindent 
Note that these results are identical to those presented at eq. (8.10) of Ref. \cite{Papadopoulos_2009} if we eliminate all dependence on his coordinate $v$ (which we have called $u$ instead). This concludes the proof of the fact that the ansatz used in Section~\ref{sec:SBHs} is the most general ansatz to construct heterotic string backgrounds consisting of supersymmetric superpositions of fundamental strings with momentum along them.

\section{First-order $\alpha'$-corrections to the fundamental rotating string solution}

\label{app:corrections}

We present in this Appendix the precise form of the first-order $\alpha'$ corrections to the small-black-ring solution presented in subsection \ref{subsec:frs}. For that, we just use eqs.~\eqref{eq:zpmzero} and plug them into eq.~\eqref{eq:solzp} to obtain the $\alpha'$-corrected solution. Following the notation of the main text, we encounter that

\begin{eqnarray}
\mathcal{Z}_{+} &=& \zpz+\alpha' \mathcal{Z}_{+}^{(1)} + \mathcal{O}(\alpha'{}^2)  \label{eq:zpringapp}  \, , \\    
\mathcal{Z}_{-}&=&\zmz+ \mathcal{O}(\alpha'{}^2)\,, \\ \label{eq:omegaringapp}
\omega &=&\omegaz+ \mathcal{O}(\alpha'{}^2) \,,
\end{eqnarray}

\indent
where

\begin{eqnarray}
\label{eq:zpringapp} 
\zpz &=&1+\tilde{q}_+ \left (\frac{y-x}{2 R^2 y }  \right )^{\frac{d-3}{2}} \, \, {}_2F{}_1 \left ( \frac{d-3}{4}, \frac{d-1}{4}; 1; 1-\frac{1}{y^2} \right )\, , \\    \label{eq:zmringapp} 
\mathcal{Z}_{-}^{(0)}&=&1+ \tilde{q}_- \left ( \frac{y-x}{2 R^2 y }  \right )^{\frac{d-3}{2}} \, \, {}_2F{}_1 \left ( \frac{d-3}{4}, \frac{d-1}{4}; 1; 1-\frac{1}{y^2} \right )\,, \\ \label{eq:omegaringapp}
\omega^{(0)} &=&\frac{(d-3) \pi \tilde{q}_-  {\mathscr W}  R^4  (y^2-1) }{\ell (x-y)^2} \left ( \frac{y-x}{2 R^2 y } \right )^{\frac{d-1}{2}}  {}_2F{}_1 \left ( \frac{d-1}{4}, \frac{d+1}{4}; 2; 1-\frac{1}{y^2} \right ) d \psi  \,,\\ 
\nonumber
\mathcal{Z}_{+}^{(1)} &=&  \frac{1}{\ell^2}C(x,y;d) \Bigg [ A_1(x,y;d) \, {}_2 F{}_1 \left ( \frac{d-3}{4}, \frac{d-1}{4};1;1-\frac{1}{y^2} \right ){}_2 F{}_1 \left ( \frac{d+1}{4}, \frac{d+3}{4};2;1-\frac{1}{y^2} \right )  \\&+& \nonumber A_2(x,y;d) {}_2 F{}_1 \left ( \frac{d-1}{4}, \frac{d+1}{4};2;1-\frac{1}{y^2} \right ) {}_2 F{}_1 \left ( \frac{d-1}{4}, \frac{d+5}{4};2;1-\frac{1}{y^2} \right ) \\ \nonumber &+& A_3(x,y;d) {}_2 F{}_1 \left ( \frac{d-3}{4}, \frac{d-1}{4};1;1-\frac{1}{y^2} \right )^2 +A_4(x,y;d) {}_2 F{}_1 \left ( \frac{d-1}{4}, \frac{d+1}{4};2;1-\frac{1}{y^2} \right )^2 \\ \label{eq:zpcorrected}  &+& A_5(x,y;d) {}_2 F{}_1 \left ( \frac{d-1}{4}, \frac{d+5}{4};2;1-\frac{1}{y^2} \right )^2 \Bigg] \, ,
\end{eqnarray}

\noindent
with the definitions, 

\begin{eqnarray}
C(x,y;d)&=&-\frac{2^{-4-d} y^{-3-d} (d-3)^2 \tilde{q}_- R^{4-2d} (y-x)^{d-2}}{1+2^{\frac{3-d}{2}} \tilde{q}_- (R^2 y)^{\frac{3-d}{2}} (y-x)^{\frac{d-3}{2}} \, \, {}_2 F{}_1 \left ( \frac{d-3}{4},\frac{d-1}{4};1;1-\frac{1}{y^2}\right )} \, , \\
A_1(x,y;d)&=& 8(d-1)\ell^2 \tilde{q}_+ x y^2(-1+y^2)\, , \\ \nonumber
A_2(x,y;d) &=& -4 (d+1) \left(2 \pi ^2 \tilde{q}_- R^2 {\mathscr W}^2 y^2 \left((d-5) x y^2-(d-3) y+2 x\right)- \right. \\ & & \left. \ell ^2 \tilde{q}_+ \left(y^2-1\right) (x-y)\right) \, ,\\
A_3(x,y;d) &=& 16 \ell^2 \tilde{q}_+ y^4 (x+y)\, , \\ \nonumber
A_4(x,y;d) &=&  4 \left(\pi ^2 \tilde{q}_- R^2 {\mathscr W}^2 y^2 \left((d-5) (d-1) x y^2-(d-5)^2 y^3-4 (d-4) y+4 x\right) \right. \\ & & \left. -\ell ^2\tilde{q}_+ \left(y^2-1\right) (x-y)\right)\, , \\
A_5 (x,y;d) &=& -(d+1)^2 (x-y) \left(\ell^2 \tilde{q}_+ \left(y^2-1\right)-4 \pi ^2 \tilde{q}_- R^2 {\mathscr W}^2 y^2\right)\,.
\end{eqnarray}

The different hypergeometric functions appearing at the corrected solution have the following behaviour (as $y \rightarrow -\infty$): 

\begin{eqnarray} \nonumber
{}_2 F {}_1 \left ( \frac{d-3}{4}, \frac{d-1}{4}; 1; 1-\frac{1}{y^2} \right ) \underset{y \rightarrow -\infty}{\sim} \vert y \vert^{d-4}\, , & &  {}_2 F {}_1 \left ( \frac{d+1}{4}, \frac{d+3}{4}; 2; 1-\frac{1}{y^2} \right ) \underset{y \rightarrow -\infty}{\sim} \vert y \vert^{d-2}\, ,\\ \nonumber {}_2 F {}_1 \left ( \frac{d-1}{4}, \frac{d+1}{4}; 2; 1-\frac{1}{y^2} \right ) \underset{y \rightarrow -\infty}{\sim} \vert y \vert^{d-4}\, , & &  {}_2 F {}_1 \left ( \frac{d-1}{4}, \frac{d+5}{4}; 2; 1-\frac{1}{y^2} \right ) \underset{y \rightarrow -\infty}{\sim} \vert y \vert^{d-2}\,,
\end{eqnarray}
\noindent
Consequently, a careful analysis reveals that

\begin{eqnarray}
\label{eq:horbzpapp}
\zp &\underset{y \rightarrow -\infty}{\sim}&  \vert y \vert^{d-4} +\alpha' \vert y \vert ^{d-2}\, , \\
\label{eq:horbzmapp}
\zm &\underset{y \rightarrow -\infty}{\sim}& \vert y \vert^{d-4}\, ,\\
\label{eq:horbomegaapp}
\omega &\underset{y \rightarrow -\infty}{\sim}& \vert y \vert^{d-4} d \psi\,.
\end{eqnarray}

\bibliographystyle{JHEP}
\bibliography{references}

\end{document}